\documentclass[12pt,a4paper]{article}

\RequirePackage{amsmath}
\RequirePackage{mathrsfs}
\RequirePackage{amsfonts}
\RequirePackage{dsfont}
\RequirePackage{braket}
\RequirePackage{tabularx}
\RequirePackage{nicefrac}
\RequirePackage{graphicx}
\RequirePackage{booktabs}
\RequirePackage{colortbl}
\RequirePackage{xcolor}
\RequirePackage[symbol]{footmisc}

\def\AEF{A.E. Faraggi}

\def\IJMP#1#2#3{{\it Int.\ J.\ Mod.\ Phys.}\/ {\bf A#1} (#2) #3}

\def\EJP#1#2#3{{\it Eur.\ Phys.\ Jour.}\/ {\bf C#1} (#2) #3}

\def\JHEP#1#2#3{{\it JHEP}\/ {\bf #1} (#2) #3}
\def\NPB#1#2#3{{\it Nucl.\ Phys.}\/ {\bf B#1} (#2) #3}
\def\PLB#1#2#3{{\it Phys.\ Lett.}\/ {\bf B#1} (#2) #3}
\def\PRD#1#2#3{{\it Phys.\ Rev.}\/ {\bf D#1} (#2) #3}
\def\PRL#1#2#3{{\it Phys.\ Rev.\ Lett.}\/ {\bf #1} (#2) #3}

\def\etal{{\it et al\/}}

\begin{document}
\begin{titlepage}
\samepage{
\setcounter{page}{1}
\rightline{}
\rightline{June 2018}

\vfill
\begin{center}
{\Large \bf{Classification\\ \medskip of Left-Right Symmetric
Heterotic String Vacua}}

\vspace{1cm}
\vfill

{\large Alon E. Faraggi$^{1}$\footnote{E-mail address: alon.faraggi@liv.ac.uk}
Glyn Harries$^{1}$\footnote{E-mail address: g.harries@liv.ac.uk} and John
Rizos$^{2}$\footnote{E-mail address: irizos@uoi.gr}}\\

\vspace{1cm}

{\it $^{1}$ Dept.\ of Mathematical Sciences, University of Liverpool, Liverpool
L69 7ZL, UK\\}

\vspace{.05in}

{\it $^{2}$ Department of Physics, University of Ioannina, GR45110 Ioannina,
Greece\\}

\vspace{.025in}
\end{center}

\vfill
\begin{abstract}
\noindent
The classification method of the free fermionic heterotic string vacua is
extended to models where the $SO(10)$ GUT symmetry is broken directly at the
string scale to the Left-Right Symmetric subgroup. 
Our method involves using a fixed set of basis vectors which are defined 
by the boundary conditions assigned to the free fermions before 
enumerating the string vacua by varying
the Generalised GSO (GGSO) projection coefficients.
It allows the derivation of algebraic expressions for the
GGSO projections for each sector that generates massless
states in the models. This enables a computerised 
analysis of the entire massless spectrum of a given choice of 
GGSO projection coefficients. 
The total number of vacua in the class of models chosen is $2^{66}
\approx 7.38 \times 10^{19}$. A statistical sampling is performed and a sample
size of $10^{11}$ vacua with the Left-Right Symmetric gauge group is
extracted. We present the results of the classification, noting that contrary 
to the previous classification of Pati-Salam models, no three generation 
exophobic models were found. The results obtained demonstrate the  
existence of three generation models with the necessary Higgs representations
needed for viable  spontaneous symmetry breaking, and with a leading top quark 
Yukawa coupling.
\end{abstract}

\smallskip}
\end{titlepage}

\section{Introduction}
\normalsize

The Standard Model of particle physics provides viable perturbative
parameterisation of all subatomic observable data.
The perturbative logarithmic evolution of the Standard Model
parameters hints that the Standard Model may persist in providing
viable parameterisation up to the Planck scale, where the
gravitational interaction  becomes of comparable strength.
Further elucidation of the Standard Model parameters therefore
necessitates the synthesis of the Standard Model with gravity.
Contemporary string theories provide consistent perturbative frameworks
to study this synthesis. They are chiral and
free of gauge and gravitational anomalies. String theories
are the only contemporary
theories that achieve this feat, and therefore provide a superior
framework to study the synthesis of the Standard Model with gravity.
Furthermore, the Standard Model matter charges strongly hint
at the realisation of $SO(10)$ Grand Unified Theory (GUT)
multiplet structure in nature.
By accommodating spinorial $SO(10)$
\textbf{16} representations in its perturbative spectrum,
the heterotic $E_8\times E_8$ string incorporates
the $SO(10)$--GUT picture \cite{hetecan}. A key prediction of the
$SO(10)$--GUT theory is that the Weinberg angle at the
unification scale is given by $\sin^2\theta_W(M_{\rm GUT})=3/8$,
which is compatible with the low energy data.

The string consistency conditions introduce additional
degrees of freedom beyond those of the Standard Model.
These may be interpreted as a number of extra spacetime bosonic
coordidates, or as a finite number of free, or interacting, fields
propogating on the string worldsheet.
String vacua are obtained by specifying these extra degrees of freedom,
subject to the string consistency conditions.
A priori the number of possibilities
is vast and there is no clear guiding principle to select among
them. The best we can accomplish at present is to extract
features of classes of compactifications and to develop the
methodolgy to discern between different classes \cite{spreview}.
The $\mathds{Z}_2\times \mathds{Z}_2$ toroidal orbifolds
represent one such class of models
that has been studied in detail \cite{z2xz2}.
This class of compactifications gives rise to an abundance of viable three
generation models with different unbroken subgroups of $SO(10)$,
and the canonical $SO(10)$--GUT prediction
$\sin^2\theta_W(M_{\rm String})=3/8$.
Among the distinct features of the class of
$\mathds{Z}_2\times \mathds{Z}_2$ orbifolds
we note the spinor--vector duality \cite{fkr,cfkr},
that generalises to other classes of string vacua \cite{panos};
the existence of exophobic vacua \cite{acfkr};
and the possibility to fix all geometric moduli by asymmetric
assignment of worldsheet boundary conditions \cite{moduli}.
The $\mathds{Z}_2\times \mathds{Z}_2$ torodial orbifolds have been
studied primarily by
using the free fermionic formulation of the heterotic--string in four
dimensions \cite{fff}.
These models
correspond to toroidal $\mathds{Z}_2\times \mathds{Z}_2$
orbifold compactifications
at special points in the moduli space with discrete
Wilson lines \cite{z2xz2}. Since the twisted matter spectrum in the
$\mathds{Z}_2\times \mathds{Z}_2$ orbifolds is independent of the moduli,
analysing these models at the free fermionic point captures
the phenomenological properties of the physical spectrum.
Deformations away from the free fermionic point are obtained
by adding worldsheet Thirring interactions \cite{Thirring1987}.
The fermionic formulation facilitates the analysis of the
spectrum and interactions, but the physical properties of this
class of string vacua are rooted in the $\mathds{Z}_2\times \mathds{Z}_2$
orbifold
structure.

Three generation free fermionic models have been constructed
since the late eighties. The early viable
constructions provided isolated examples with
$SU(5)\times U(1)$ (FSU5) \cite{fsu5},
$SO(6)\times SO(4)$ (PS) \cite{alr},
$SU(3)\times SU(2)\times U(1)^2$ (SLM) \cite{slm} and
$SU(3)\times U(1)\times SU(2)^2$ (LRS) \cite{lrs},
unbroken subgroups of $SO(10)$.
All with the canonical $SO(10)$ embedding of the weak hypercharge, yielding
the GUT prediction for the Weinberg angle $\sin^2\theta_W=3/8$ at
$M_{\rm String}$. The case of the $SU(4)\times SU(2)_L\times U(1)_R$ (SU421)
$SO(10)$ subgroup was shown not to produce viable models
\cite{su421, 421}. Over the past two decades systematic methods
to classify large spaces of free fermionic models were developed,
\cite{gkr, fknr, fkr, acfkr, su62,frs,hasan}, culminating in the
classification of Standard-like Models \cite{SLM}.
The initial application was for type II superstring vacua \cite{gkr},
and was extended to the classification of heterotic--string vacua
with unbroken $SO(10)$ symmetry in refs. \cite{fknr,fkr}. This
led to the discovery of spinor--vector duaility in the space
$\mathds{Z}_2\times \mathds{Z}_2$ orbifolds with $(2,0)$ worldsheet
supersymmetry
\cite{fkr, cfkr}. The classification method provides an
effective trawling algorithm to construct string models with
specific phenomenological properties. Examples include:
the construction of exophobic heterotic--string vacua \cite{acfkr};
the construction of heterotic--string vacua with $SU(6)\times SU(2)$
GUT group \cite{su62}; the construction of string vacuum that allows a family
universal extra vector boson with $E_6$ embedding to remain
unbroken down to low scales \cite{frzprime}.

In this paper we extend the classification methodology
to the class of free fermionic heterotic--string vacua
in which the $SO(10)$ gauge group is broken to the
left--right symmetric
$SU(3)\times U(1)_{B-L}\times SU(2)_L\times SU(2)_R$ extension
of the Standard Model gauge group. From a phenomenological
low energy point of view, this extension of the Standard Model
is highly motivated \cite{lrsftmodel} and dates back
to the mid--70s, when the gauge and matter
structure of the Standard Model crystallised.
The left--right symmetry naturally explains the
generation of parity violation in nature, via the
spontaneous symmetry breaking of $SU(2)_R$. It represents a
rather minimal extension of the Standard Model, that avoids
the tight constraints imposed
from proton lifetime limits
on the scale of other extensions.
Thus, the spacetime vector bosons
of the enhanced symmetry may exist below, say, 100TeV
and appear in future experiments. From the point of view
of the string model building this class of models introduces novel
characteristics that were highlighted in ref. \cite{lrs}. The
basic contrast from the FSU5, PS and SLM models is that the
LRS models of ref. \cite{lrs} do not possess the $E_6$ embedding
of the Standard Model states. This results in left-right
symmetric models that do not have an anomalous $U(1)$ symmetry,
which appears in the others cases. In that respect the LRS models
of ref. \cite{lrs} possess a similar structure to the
SU421 models of ref. \cite{su421, 421} that were shown not to
admit viable three generation models. These properties of the
LRS free fermionic models may be traced to the absence of an $x$--map
\cite{xmap} in the LRS models of ref. \cite{lrs}. By contrast the
LRS models that we construct herein do possess an $x$--map, which
represents a vital step in the classification methodology.
In that respect the models that we analyse in this paper
are distinct from those of ref. \cite{lrs}.

Our paper is organised as follows: in section \ref{lrsffm} a brief 
introduction to the construction of free fermionic models is given. 
The set of basis vectors used to generate $SO(10)$ models are presented, 
before outlining the construction of the LRS models in the ensuing discussion. 
Section \ref{stringspec} presents details of the string spectrum, 
such as the untwisted gauge symmetry of the LRS models and enumeration of
the sectors that can enhance it. 
Section \ref{TMS} provides a complete description of the twisted matter 
spectrum. This includes the observable, exotic and hidden matter sectors. 
Section \ref{CRAL} details the results of the classification and contains 
analysis of the data while providing comparisons to earlier classifications 
\cite{acfkr, frs, SLM}. In section \ref{notable} we give the GGSO projection
coefficients of a phenomenologically viable LRS heterotic--string model. 
In our examplery model all enhancing vector bosons are projected out. 
It contains three chiral generations, no chiral exotics as well as 
heavy and light Higgs multiplets, required for viable 
spontaneous symmetry breaking and fermion masses. 
Section \ref{conclusion} concludes the paper.

\section{Left Right Symmetric Free Fermionic Models}\label{lrsffm}

This paper concerns the extension of the free fermionic classification method,
utilised in \cite{fknr, fkr, acfkr, frs, SLM}, to vacua which possess the
Left-Right Symmetric (LRS) subgroup of $SO(10)$. The free fermionic models
correspond to $\mathds{Z}_2 \times \mathds{Z}_2$ orbifold compactifications
with $\mathcal{N} = (2,0)$ superconformal worldsheet supersymmetry and discrete
Wilson lines. The formulation of the free fermions occurs at an extented
symmetry point in the moduli space where the compactified directions are
interpreted as two dimensional fermionic degrees of freedom which propagate on
the string worldsheet.

The free fermionic formulation provides a set of rules which enables extraction
of the physical states in a string model and provides a straightforward
approach to studying the phenomenological properties of the string vacua.
The models are constructed by defining a set of basis vectors and the 
Generalised Gliozzi-Scherk-Olive (GGSO) projection coefficients of 
the one-loop partition function. The details are outlined in
the following section.

The breaking of the $SO(10)$ GUT symmetry occurs directly at the string scale.
All the models which are classified possess $\mathcal{N}=1$ spacetime
supersymmetry and preserve the $SO(10)$ embedding of the weak hypercharge. The
unbroken subgroup of $SO(10)$ in the low energy effective field theory
considered here is $SU(3)_C\times U(1)_C \times SU(2)_L \times SU(2)_R$. The
matter states which give rise to the Standard Model fermionic representations
are found in the spinorial \textbf{16} representation of $SO(10)$ 
decomposed under the
unbroken $SO(10)$ subgroup. Similarly, the SM light Higgs states occur from the
vectorial \textbf{10} representation of $SO(10)$.

\subsection{The Free Fermionic Formulation}
The notable features of the free fermionic formulation used in model building
and classification will be briefly outlined. A more detailed discussion of
these features can be found in reference \cite{fff}.

The free fermionic formulation of string theory is directly formulated in four
space-time dimensions, whereby the extra degrees of freedom found in string
theories are interpreted as free fermions propagating on the two dimensional
string worldsheet. The approach considered here utilises the four dimensional
heterotic string in the light cone gauge, meaning there are 20 left moving and
44 right moving free fermions introduced to account for all the extra degrees of
freedom. In the standard notation the left movers are represented by
$\psi^{\mu}_{1,2} \; , \; \chi^{1,\ldots,6} \; , \; y^{1,\ldots,6}\; , \;
w^{1,\ldots,6}$ and the right movers by $\overline{y}^{1,\ldots,6}\; ,
\;\overline{w}^{1,\ldots,6}\; , \;\overline{\psi}^{1,\ldots,5}\; ,
\;\overline{\eta}^{1,2,3}\; , \; \overline{\phi}^{1,\ldots,8}$.

When these fermions are parallel transported around the two noncontractible
loops of the one-loop partition function, they obtain a non-trivial
phase\footnote[1]{In the common nomenclature, these phases 
are also referred to as
`boundary conditions' of the free fermions.}. These phases can be either
periodic, anti-periodic or complex, denoted by 0,1 and $\pm\frac{1}{2}$
respectively. The boundary conditions of the fermions are specified in
64-dimensional vectors called `basis vectors' which are given in the form
\begin{equation*}
v_i = \{ \alpha_i(f_1), \ldots, \alpha_i(f_{20}) \; | \;
\alpha_i(\overline{f}_1) , \ldots, \alpha_i(\overline{f}_{44}) \},
\end{equation*}
where the boundary condition $\alpha$ is defined as the transformation property
for a fermion $f$. Accordingly,
\begin{equation*}
f_j \rightarrow -e^{i\pi \alpha_i (f_j)} f_j, \qquad j = 1,\ldots,64.
\end{equation*}

Each model is specified by a set of basis vectors $v_1, \ldots, v_N$, which must
satisfy modular invariance constraints. The basis vectors of the model span a
space $\Xi$, which consists of $2^{N+1}$ sectors. Each sector is formed as a
linear combination of the basis vectors and is given by
\begin{equation}
\xi = \sum^{N}_{i=1} m_j v_i \qquad m_j = 0,1,\ldots,N_j-1,
\end{equation}
where $N_j\cdot v_j = 0 \mod 2$. The string states in each sector, denoted by
$\ket{S_{\xi}}$, must also conform to modular invariance constraints. This is
imposed on the string states in the form of the one-loop GGSO projections via
the equation,
\begin{equation}
e^{i\pi v_i\cdot F_{\xi}}\ket{S_{\xi}} = \delta_{\xi}\; C\binom{\xi}{v_i}^*
\ket{S_{\xi}},
\end{equation}
where $F_{\xi}$ is the fermion number operator, $\delta_{\xi} = \pm 1$ is the
space-time spin statistics index and $C\binom{\xi}{v_i} = \pm 1\; ;\; \pm
\frac{1}{2}$ is the GGSO projection coefficient. By varying the choice of the
GGSO coefficients, distinct vacua of the string model are obtained.

Summarising, a model is constructed by using a set of 
basis vectors $v_i$ and a
set of distinct GGSO projection coefficients $C\binom{v_i}{v_j}$, 
with $i>j$, of which there are $2^{N(N-1)/2}$.

\subsection{$SO(10)$ Models}
In order to build the Left-Right Symmetric models that are studied in this
paper, a set of thirteen basis vectors are used. The first twelve basis vectors
considered generate $SO(10)$ models and are common in the previous publications
\cite{acfkr, frs, SLM, 421}. These basis vectors are also included in the basis
of the LRS models discussed here and are defined as:

\begin{eqnarray}
v_1={\mathds{1}}&=&\{\psi^\mu,\
\chi^{1,\dots,6},y^{1,\dots,6}, \omega^{1,\dots,6}| \nonumber\\
& & ~~~\overline{y}^{1,\dots,6},\overline{\omega}^{1,\dots,6},
\overline{\eta}^{1,2,3},
\overline{\psi}^{1,\dots,5},\overline{\phi}^{1,\dots,8}\},\nonumber\\
v_2=S&=&\{{\psi^\mu},\chi^{1,\dots,6}\},\nonumber\\
v_{2+i}={e_i}&=&\{y^{i},\omega^{i}\; | \; \overline{y}^i,\overline{\omega}^i\},
\
i=1,\dots,6,\nonumber\\
v_{9}={b_1}&=&\{\chi^{34},\chi^{56},y^{34},y^{56}\; | \; \overline{y}^{34},
\overline{y}^{56},\overline{\eta}^1,\overline{\psi}^{1,\dots,5}\},\label{basis}\\
v_{10}={b_2}&=&\{\chi^{12},\chi^{56},y^{12},y^{56}\; | \; \overline{y}^{12},
\overline{y}^{56},\overline{\eta}^2,\overline{\psi}^{1,\dots,5}\},\nonumber\\
v_{11}=z_1&=&\{\overline{\phi}^{1,\dots,4}\},\nonumber\\
v_{12}=z_2&=&\{\overline{\phi}^{5,\dots,8}\},
\nonumber
\end{eqnarray}
where $i = 1,\ldots,6$ and the fermions which appear in the basis vectors have
periodic (Ramond) boundary conditions, whereas those not included have
antiperiodic (Neveu-Schwarz) boundary conditions.

The basis vector $\mathds{1}$ is required by the rules set out in the papers
listed in reference \cite{fff} and generates a model with an $SO(44)$ gauge
group from the Neveu-Schwarz (NS) sector. Addition of the $S$ basis vector
generates $\mathcal{N} = 4$ space-time supersymmetry and leaves the gauge group
intact. The $e_i$ vectors break the gauge group to $SO(32)\times U(1)^6$ but
preserve the $\mathcal{N}=4$ supersymmetry. These vectors correspond to all the
possible internal symmetric shifts of the six internal bosonic coordinates.
Addition of the vectors $b_1$ and $b_2$ corresponds to $\mathbb{Z}_2 \times
\mathbb{Z}_2$ orbifold twists and breaks the space-time supersymmetry firstly
to $\mathcal{N}=2$ and subsequently to $\mathcal{N}=1$. They also break the
$U(1)^6$ gauge symmetry, therefore reducing the rank of the gauge group, while
simultaneously decomposing the $SO(32)$ to $SO(10)\times U(1)^3 \times SO(16)$.
Addition of the basis vectors $z_1$ and $z_2$ then break the hidden $SO(16)$
gauge group, generated by the fermions $\overline{\phi}^{1,\ldots,8}$, to
$SO(8)\times SO(8)$.
The untwisted vector bosons present due to this choice of basis vectors
generate the gauge group $SO(10) \times U(1)^3 \times SO(8)^2$ in the adjoint
representation.

\subsection{Left-Right Symmetric Models}
Previous constructions of free fermionic LRS models used two or more basis
vectors to break the observable gauge group. Firstly, one basis vector with
either the assignment $\overline{\psi}^{1,2,3} = 1$ (as in \cite{lrs}), or
equivalently $\overline{\psi}^{4,5} = 1$ (as in \cite{acfkr}), is used to
obtain the $SO(6)\times SO(4)$ Pati-Salam gauge group and a second basis vector
with the assignment $\overline{\psi}^{1,2,3} = \pm\frac{1}{2}$ breaks the
Pati-Salam gauge group to the LRS.

However, the model under consideration here uses only one additional basis
vector, given by
\begin{equation}
\alpha = \{ \overline{\psi}^{1,2,3} = \frac{1}{2} \; , \;
\overline{\eta}^{1,2,3} = \frac{1}{2}\; , \; \overline{\phi}^{1,\ldots,6} =
\frac{1}{2}\; , \; \overline{\phi}^7 \},
\end{equation}
where the restriction that the phase on the complex right-moving fermions is
positive is made, \textit{i.e} $\overline{\psi}^{1,2,3}=+\frac{1}{2}$. The
assignment of $\overline{\eta}^{1,2,3} = +\frac{1}{2}$ is made due to the
contraint that $b_{j} \cdot \alpha = 0\mod 1$, where $j=1,2,3$, must be true in
order to satisfy modular invariance.

It should be noted that while the assignments on the fermions
$\overline{\psi}^{1,2,3},\overline{\eta}^{1,2,3}$ must be as above, this choice
of $\alpha$ is not unique due to possible variations of assignments for 
the fermions $\overline{\phi}^{1,\ldots,8}$. However, in this paper only
models with the $\alpha$ defined above are considered.

With this choice of basis vectors, we note two sectors which 
are combinations of the basis vectors and facilitate the classification
and presentation of the physical spectrum. The first 
is the composite vector defined as `$b_3$' which is given by
\begin{equation}
\begin{split}
b_3 & = \mathds{1} + S + \sum^{6}_{i=1} e_i + b_1 + b_2 + z_1 + z_2\\
& = \{ \chi^{12} , \chi^{34} , y^{12} , y^{34} , \; | \; \overline{y}^{12} ,
\overline{y}^{34} , \overline{\psi}^{1,\ldots,5} , \overline{\eta}^3 \}.
\end{split}
\end{equation}
This combination of basis vectors corresponds to the third twisted 
plane of the $\mathds{Z}_2\times \mathds{Z}_2$ 
orbifold, where the first two are 
related to $b_1$ and $b_2$, respectively. 
The second is given by the linear combination denoted by `$x$', 
given by
\begin{equation}\label{x}
\begin{split}
x & = \mathds{1} + S + \sum^{6}_{i = 1} e_i + z_1 + z_2 \\
& = \{ \overline{\psi}^{1,2,3,4,5} \; , \; \overline{\eta}^{1,2,3} \}.
\end{split}
\end{equation}
This linear combination produces the spinorial 128 multiplet
in the 248 adjoint representation of the observable $E_8$, generated
by the subset $\{1, S, x, z_1+z_2\}$ of the basis set (\ref{basis}).
It generates the so--called $x$--map \cite{xmap} that exchanges 
spinorial and vectorial representations from the twisted sectors 
$B_j$, to be defined below, and $B_j+x$, respectively. We remark
that this linear combination is not generated in the LRS models 
of ref. \cite{lrs} and therefore the models presented there 
do not admit the $x$--map. This is an important distinction between
the models considered here and those of ref. \cite{lrs}. We note that the
$x$--map is crucial in our classification method as the sectors
$B_j+x$ are those that give rise to the Standard Model electroweak
doublets. 
Therefore, the basis of the models considered consists of the basis vectors $\{
\mathds{1},S,e_1,e_2,e_3,e_4,e_5,e_6,b_1,b_2,z_1,z_2,\alpha\}$ with two notable
linear combinations $\{b_3,x\}$.

\subsection{GGSO Projections}
Now that the basis has been specified, the next components of the model which
need defining are the GGSO projection coefficients $C\binom{v_i}{v_j}$ which
are necessary in order to completely describe the one-loop partition function.

The GGSO coefficients span a $13\times 13$ matrix. The lower triangle of the
matrix containing 78 coefficients is fixed by the corresponding 78 coefficients
in the upper triangle by modular invaraince constraints. In addition, the
phases on the diagonal are also fixed by modular invariance.
Accordingly,
\begin{equation}
\begin{split}
& C\binom{e_i}{e_i} = -C\binom{e_i}{\mathds{1}}\qquad i=1,\ldots, 6\\
& C\binom{b_k}{b_k} = C\binom{b_k}{\mathds{1}}\qquad k=1,2\\
& C\binom{z_k}{z_k} = C\binom{z_k}{\mathds{1}}\qquad k=1,2\\
& C\binom{\alpha}{\alpha} = C\binom{\alpha}{\mathds{1}}.
\end{split}
\end{equation}
To ensure $\mathcal{N} = 1$ supersymmetry, 
without loss of generality, the following coefficients
are fixed
\begin{equation}
C\binom{\mathds{1}}{\mathds{1}}=C\binom{S}{\mathds{1}} = C\binom{S}{S} =
C\binom{S}{e_i} = C\binom{S}{b_k} = C\binom{S}{z_k} = C\binom{S}{\alpha} = -1,
\end{equation}
where $i=1,\ldots,6$ and $k=1,2$. We are therefore left with 66 independent
coefficients, which generates $2^{66} \approx 7.38\times 10^{19}$ distinct
string vacua.

It should be noted that all the phases are real and take the discrete values
$\pm 1$ except for the phase $C\binom{\mathds{1}}{\alpha}$ which takes values
$\pm i$ due to the fact that $\mathds{1}\cdot\alpha = -7$.

\section{String Spectrum}\label{stringspec}
Adapting the methodology of previous cases 
\cite{acfkr,421, frs, SLM}, 
the sectors which can contribute massless states are enumerated and the
corresponding algebraic conditions for the GGSO projections are derived for
each sector.

Spacetime vector bosons that arise from the 
the untwisted sector,  generate the
$SO(10)$ symmetry and its unbroken subgroups. 
There are further sectors in these
models that can give rise to additional physical 
spacetime vector bosons, 
which enhance the untwisted gauge symmetry.
Furthermore, if the additional spacetime 
vector bosons are charged with respect to the Cartan generators of 
the $SO(10)$ GUT symmetry, the unbroken $SO(10)$ subgroup is enhanced. 
Thus, a pivotal requirement in the construction is that the
additional spacetime vector bosons are projected out. 

The twisted sectors produce matter multiplets which possess $\mathcal{N} = 1$
supersymmetry and can be grouped depending on which $SO(10)$ subgroup they
leave unbroken. Sectors which contain the $\alpha$ basis vector in the linear
combination break the $SO(10)$ symmetry to the LRS and gives rise to exotic
states. If the linear combination contains $2\alpha$ then the $SO(10)$ gauge
group is broken to the Pati-Salam $SO(6)\times SO(4)$ gauge group and also
contains exotics. As $\alpha$ is the only $SO(10)$ breaking basis vector, all
the remaining sectors which, a priori, do not include $\alpha$ in the linear
combination do not break the $SO(10)$ symmetry.

The sectors in a model can be catergorised according to the left and right
moving vacuum. The physical states satisfy the Virasoro condition, defined as
\begin{equation}
M_L^2 = -\frac{1}{2} + \frac{\xi_L \cdot \xi_L}{8} + N_L = -1 + \frac{\xi_R
\cdot \xi_R}{8} + N_R = M_R^2
\end{equation}
where $N_L$ and $N_R$ are the sums over the left and right moving oscillators
respectively. Sectors that have the products $\xi_L\cdot \xi_L = 0$ and
$\xi_R\cdot\xi_R = 0,4,6,8$ can produce spacetime vector bosons, which
determine the gauge symmetry in a given vacuum. Sectors where the products are
$\xi_L\cdot \xi_L = 4$ and $\xi_R\cdot\xi_R = 4,6,8$ produce matter states
which are outlined in section 4. All the models considered preserve
$\mathcal{N} = 1$ spacetime supersymmetry, which is generated by the basis
vector $S$ where the products are $(S_L \cdot S_L\; ; \; S_R \cdot S_R)
= (4;0)$.

\subsection{The Gauge Symmetry}
Vector bosons from the untwisted sector correspond to generators of the
following observable and hidden gauge group symmetries
\begin{eqnarray}
\text{Observable}&:& SU(3)_C \times U(1)_C \times SU(2)_L \times SU(2)_R
\times U(1)_1 \times U(1)_2 \times U(1)_3~~~~~~\label{observable}\\
\text{Hidden}&:&  SU(4) \times U(1)_4 \times SU(2)_5 \times U(1)_5 \times
U(1)_7 \times U(1)_8\label{hidden}
\end{eqnarray}
and the weak hypercharge is given 
by\footnote[2]{It should be noted that $U(1)_C =
\frac{3}{2}U(1)_{B-L}$ and $U(1)_L = 2U(1)_{T_{3_R}}$}
\begin{equation}
U(1)_Y = \frac{1}{3}U(1)_C + \frac{1}{2}U(1)_L. 
\end{equation}

Depending on the choice of GGSO projection coefficients, additional space-time
vector bosons may arise from the following 26 sectors
\begin{equation}\label{G}
\textbf{G} =
\begin{Bmatrix}
x & z_1 & z_2 & z_1 + z_2 \\
&&& \\
z_1 + 2\alpha & z_1 + z_2 + 2\alpha & 2\alpha + x & z_2 + 2\alpha + x \\
z_1 + 2\alpha + x & z_1 + z_2 + 2\alpha + x & & \\
&&& \\
\alpha & 3\alpha & z_1 + \alpha & z_1 + 3\alpha \\
z_2 + \alpha & z_2 + 3\alpha & z_1 + z_2 + \alpha & z_1 + z_2 + 3\alpha \\
\alpha + x & 3\alpha + x & z_1 + \alpha + x & z_1 + 3\alpha + x \\
z_2 + \alpha + x & z_2 + 3\alpha + x & z_1 + z_2 + \alpha + x & z_1 + z_2 +
3\alpha + x
\end{Bmatrix},
\end{equation}
where $x$ is defined in equation (\ref{x}). The sectors in (\ref{G}) have been
organised such that the sectors which do not break the $SO(10)$ symmetry are on
row 1; rows 2-3 break the $SO(10)$ symmetry to the Pati-Salam $SO(6) \times
SO(4)$ gauge group and finally rows 4-7 break the $SO(10)$ symmetry to the LRS
$SU(3)\times U(1) \times SU(2)\times SU(2)$ gauge group.

We remark
that any projections on sectors containing $3\alpha$ can
be inferred from the projections made on the corresponding sector which 
contains only
$\alpha$. Therefore, in the following analysis these sectors will not
be discussed in detail.

If any of the gauge bosons from the sectors in eq. (\ref{G})
survive the projections, the untwisted gauge
symmetry is enhanced. We restrict the classification analysis to vacua with no
enhancements, meaning that the gauge symmetry of all the vacua classified is
identical. In the classification method the GGSO projection coefficients of the
26 sectors listed above were derived and expressed in an analytic form so that
a computer code can easily detect if a particular vacua is enhanced. Of the
vacua that were scanned in the classification, approximately $29.1\%$
contained extra vector bosons and were therefore enhanced.

\section{The Twisted Matter Spectrum}\label{TMS}
\subsection{General Remarks}
In the table below, the hypercharge and electromagnetic charge have been
normalised according to the equations
\begin{subequations}
\begin{equation}\label{Y}
 Y = \frac{1}{3}( Q_1 + Q_2 + Q_3) + \frac{1}{2} (Q_4 + Q_5) \\
\end{equation}
\begin{equation}\label{Qem}
 Q_{em} = Y + \frac{1}{2}( Q_4 - Q_5 )
\end{equation}
\end{subequations}
In these equations, the $U(1)$ charges $Q_{1,\ldots,5}$ are the $U(1)$ charges
generated by the fermions $\overline{\psi}^{1,\ldots,5}$ respectively and are
calculated according to the equation
\begin{equation}
Q(f) = \frac{1}{2}\alpha(f) + F(f)
\end{equation}
where $\alpha(f)$ is the boundary condition of the fermion in the sector and
$F(f)$ is the fermion number given by
\begin{subequations}
\begin{equation}
F (f)=\begin{cases}+1&\text{for}\; f\\ -1&\text{for}\; f^* \end{cases}
\end{equation}
for fermionic oscillators and their complex conjugates, and
\begin{equation}
\begin{split}
& F \ket{+}_R = 0 \\
& F \ket{-}_R = -1
\end{split}
\end{equation}
\end{subequations}
for the degenerate Ramond vacua where $\ket{+}_R = \ket{0}$ is a degenerated
vacuum with no oscillator and $\ket{-}_R = f_0^{\dagger}\ket{0}$ is the
degenerated vacua with one zero mode oscillator.

The table below outlines the electromagnetic charges, and the charges under the
electroweak $SU(2)\times U(1)$ Cartan generators, of the states which are
contained in the observable LRS chiral matter representations:

\begin{center}
\begin{tabular}{|c|c|c|c|c|}
\hline
 Representation & $\overline{\psi}^{1,2,3}$ &
$\overline{\psi}^{4,5}$ & $Y$ & $Q_{em}$ \\
\hline \hline
$\left( \textbf{3} , +\nicefrac{1}{2} \;  , \textbf{2},\textbf{1} \right)$ &
($+,+,-$)& $(+,-)$ & 1/6 & 2/3 , -1/3\\
\hline
& $(+,+,-)$& ($+,+$)& 2/3 & 2/3 \\
$\left(\textbf{3} , +\nicefrac{1}{2} \; , \textbf{1} , \textbf{2}\right)$ &
($+,+,-$)& $(-,-)$ & -1/3 & -1/3 \\
\hline
$\left(\overline{\textbf{3}} , -\nicefrac{1}{2} \; , \textbf{2} ,
\textbf{1}\right)$& ($+,-,-$)& $(+,-)$ & -1/6 & 1/3 , -2/3\\
\hline
& $(+,-,-)$& $(+,+)$ & 1/3 & 1/3 \\
$\left( \overline{\textbf{3}} , -\nicefrac{1}{2} \; , \textbf{1},\textbf{2}
\right)$ & $(+,-,-)$& $(-,-)$ & -2/3 & -2/3 \\
\hline
$\left( \textbf{1} , +\nicefrac{3}{2} \; , \textbf{2}, \textbf{1} \right)$ &
$(+,+,+)$ & $(+,-)$ & 1/2 & 1 , 0\\
\hline
& $(+,+,+)$ & $(+,+)$ & 1 & 1\\
$\left( \textbf{1} , +\nicefrac{3}{2} \; , \textbf{1} , \textbf{2} \right)$ &
$(+,+,+)$ & $(-,-)$ & 0 & 0\\
\hline
$\left( \textbf{1} , -\nicefrac{3}{2} \; , \textbf{2} , \textbf{1} \right)$ &
$(-,-,-)$ & $(+,-)$ & -1/2 & 0 , -1\\
\hline
& $(-,-,-)$ & $(+,+)$ & 0 & 0\\
$\left( \textbf{1} , -\nicefrac{3}{2} \; , \textbf{1} , \textbf{2} \right)$ &
$(-,-,-)$ & $(-,-)$ & -1 & -1 \\
\hline
\end{tabular}
\end{center}
where the representation is decomposed as $SU(3)_C \times U(1)_C \times SU(2)_L
\times SU(2)_R$. The notation `$+$' above denotes a state of the degenerated
Ramond vacua with no oscillator, \textit{i.e} a state with a fermion number $F=
0$, whereas the notation `$-$' denotes a state of the degenerated Ramond vacua
with a zero mode oscillator and therefore a state where $F= -1$. The values for
$Y$ and $Q_{em}$ are calculated using equations (\ref{Y}) and (\ref{Qem})
respectively.

It is when these representations are decomposed under the SM gauge group
$SU(3)_C \times SU(2)_L \times U(1)_Y$ that we get the particle states of the
Standard Model. The leptons and quarks are realised by the following
representations
\begin{subequations}
\begin{equation}
Q_L^i = (\textbf{3},\textbf{2},\textbf{1})_{\frac{1}{6}} = \binom{u}{d}^i,
\end{equation}
\begin{equation}
Q_R^i = (\overline{\textbf{3}},
\textbf{1},\textbf{2})_{\frac{1}{3}, -\frac{2}{3}} = \binom{d^c}{u^c}^i,
\end{equation}
\begin{equation}
L_L^i = (\textbf{1},\textbf{2},\textbf{1})_{-\frac{1}{2}} = \binom{\nu}{e}^i,
\end{equation}
\begin{equation}
L_R^i = (\textbf{1},\textbf{1},\textbf{2})_{1,0} = \binom{e^c}{\nu^c}^i,
\end{equation}
\begin{equation}
h = (\textbf{1},\textbf{2},\textbf{2})_0 = \begin{pmatrix}h_+^u & h_0^d \\ h^u_0
& h^d_- \end{pmatrix}
\end{equation}
\end{subequations}
where $h^u$ and $h^d$ are the low energy supersymmetric superfields associated
with the Minimally Supersymmetric Standard Model (MSSM).

\subsection{The Observable Matter Sectors}\label{observablematter}
The chiral matter spectrum is obtained from the twisted sectors, which are as
follows
\begin{eqnarray}
B_{pqrs}^{(1)} &=& S + b_1 + pe_3 + qe_4 + re_5 + se_6
\nonumber \\ 
&=& \{\psi^{\mu},\chi^{1,2},
(1-p)y^3\bar{y}^3,pw^3\bar{w}^3,(1-q)y^4\bar{y}^4,qw^4\bar{w}^4,
 \\
& & ~~~ (1-r)y^5\bar{y}^5,rw^5\bar{w}^5,(1-s)y^6\bar{y}^6,
sw^6\bar{w}^6,\bar{\eta}^{1},\bar{\psi}^{1,\ldots,5}\}
\nonumber \\
B_{pqrs}^{(2)} &=&  S + b_2 + pe_1 + qe_2 + re_5 + se_6\nonumber \\
B_{pqrs}^{(3)} &=&  S + b_3 + pe_1 + qe_2 + re_3 + se_4\nonumber
\end{eqnarray}
where $p,q,r,s = 0,1$ and $b_3 = b_1 + b_2 + x$. These 48 sectors contain the
$\textbf{16}$ and $\overline{\textbf{16}}$ spinorial representations of the
$SO(10)$ observable gauge group decomposed under 
$SU(3)_C \times U(1)_C \times SU(2)_L \times
SU(2)_R$ as
\begin{equation*}
\begin{split}
& \textbf{16} = (\textbf{3} , {+\textstyle\frac{1}{2}}, 
\textbf{2} , \textbf{1}) +
(\overline{\textbf{3}} , {-\textstyle\frac{1}{2}}, 
\textbf{1} , \textbf{2}) + (\textbf{1}
, {-\textstyle\frac{3}{2}}, \textbf{2} , \textbf{1}) + 
(\textbf{1} , {+\textstyle\frac{3}{2}} , \textbf{1} ,
\textbf{2}), \\
& \overline{\textbf{16}} = (\overline{\textbf{3}}, 
{-\textstyle\frac{1}{2}} , \textbf{2} ,
\textbf{1}) + (\textbf{3},{+\textstyle\frac{1}{2}}  , \textbf{1} ,
\textbf{2}) + (\textbf{1}, {+\textstyle\frac{3}{2}} , \textbf{2} ,
\textbf{1}) + (\textbf{1}, {-\textstyle\frac{3}{2}} , \textbf{1} ,
\textbf{2}).
\end{split}
\end{equation*}
We remark that in this construction, each of the sectors $B^{(i)}_{pqrs}$ with 
$i=1,2,3$,
can contribute at most a single multiplet to the physical spectrum.
The integers $\{pqrs\}$ essentially label the sixteen fixed points 
of the $i^{th}$ twisted plane. For this reason we can interchange the 
identification of the $\{pqrs\}$--sectors with states in the physical
spectrum, {\it i.e.} the spectrum of states that survive the GGSO 
projections. The power of the formalism is that all the states
producing sectors can be expressed in a similar fashion. 

In addition to the twisted matter spectrum, there are vector-like states which
contribute to the observable matter spectrum. These states arise from the
sectors
\begin{eqnarray}
B_{pqrs}^{(4)} &=& S + b_1 + pe_3 + qe_4 + re_5 + se_6 + x
\nonumber \\ 
&=& \{\psi^{\mu},\chi^{1,2},(1-p)y^3\bar{y}^3,pw^3\bar{w}^3,
(1-q)y^4\bar{y}^4,qw^4\bar{w}^4,
 \\
& & ~~~~~~~~ (1-r)y^5\bar{y}^5,rw^5\bar{w}^5,
(1-s)y^6\bar{y}^6,sw^6\bar{w}^6,\bar{\eta}^{2,3}\}
\nonumber \\
B_{pqrs}^{(5)} &=&  S + b_2 + pe_1 + qe_2 + re_5 + se_6 + x\nonumber \\
B_{pqrs}^{(6)} &=&  S + b_3 + pe_1 + qe_2 + re_3 + se_4 + x\nonumber
\end{eqnarray}
which have four periodic right-moving complex fermions. Massless states can be
obtained by acting on the vacuum with a Neveu-Schwarz right-moving fermionic
oscillator. If the oscillator is from either the fermions
$\overline{\psi}^{1,\ldots,5}$ or their complex conjugates
$\overline{\psi}^{*1,\ldots,5}$ then these sectors give rise to the vectorial
$\textbf{10}$ representation of $SO(10)$ decomposed under 
$SU(3)_C \times U(1)_C \times
SU(2)_L \times SU(2)_R$ as
\begin{equation*}
\textbf{10} = (\textbf{3} ,{-\textstyle{1}}, \textbf{1} , 
\textbf{1}) + (\overline{\textbf{3}} ,{+\textstyle{1}},
\textbf{1} , \textbf{1}) + (\textbf{1} , \textstyle{0}, 
\textbf{2} , \textbf{2}), 
\end{equation*}
where the first and second representations are generated by the fermions
$\{\overline{\psi}^{1,2,3}\}$ and $\{\overline{\psi}^{* 1,2,3}\}$ respectively
and the final representation is generated by the fermions
$\{\overline{\psi}^{4,5}\}$ and $\{\overline{\psi}^{* 4,5}\}$. It can be seen
that the first two representations are colour triplets, usually referred to as
leptoquarks in the literature, which mediate proton decay via dimension five
operators. Therefore, these states must be either sufficiently heavy so as to
agree with the current proton lifetime of $\geq 10^{33}$ years \cite{proton} or
must be projected out of the string spectrum by the GGSO projections. This is a
constraint which is considered when the classification is performed.
The representation $(\textbf{1},\textstyle{0},\textbf{2},\textbf{2})$ 
give rise to the light Standard Model Higgs.

The remaining right-moving complex fermions can give rise to states which are
singlets under the observable gauge group but form the following
representations
\begin{itemize}
\item
$\{\overline{\eta^{i}}\} \ket{R}^{(4,5,6)}_{pqrs}$ or
$\{\overline{\eta}^{*i}\}\ket{R}^{(4,5,6)}_{pqrs}$ , $i = 1,2,3,$ where
$\ket{R}^{(4,5,6)}_{pqrs}$ is the degenerated Ramond vacuum of the sectors
$B^{(4,5,6)}_{pqrs}$ respectively. These states transform as vector-like
representations of the $U(1)_i$'s.
\item
$\{\overline{\phi}^{1,\ldots,4}\} \ket{R}^{(4,5,6)}_{pqrs}$ or
$\{\overline{\phi}^{*1,\ldots,4}\}\ket{R}^{(4,5,6)}_{pqrs}$. These states
transform as vector-like representations of the $SU(4)\times U(1)_4$ gauge
group.
\item
$\{\overline{\phi}^{5,6}\} \ket{R}^{(4,5,6)}_{pqrs}$ or
$\{\overline{\phi}^{*5,6}\}\ket{R}^{(4,5,6)}_{pqrs}$. These states transform
as vector-like representations of the $SU(2)_5\times U(1)_5$ gauge group.
\item
$\{\overline{\phi}^{7,8}\} \ket{R}^{(4,5,6)}_{pqrs}$ or
$\{\overline{\phi}^{*7,8}\}\ket{R}^{(4,5,6)}_{pqrs}$. These states transform
as vector-like representations of the $U(1)_7$ and $U(1)_8$ gauge groups
respectively.
\end{itemize}

\subsubsection{Chirality Operators}
In order to calculate the number of families of a model, the number of chiral
$\textbf{16}$ and $\overline{\textbf{16}}$ representations of $SO(10)$
decomposed under the LRS gauge group have to be counted. 
In these models families and anti-families 
are formed from the following representations
\begin{equation}
\begin{split}
 \textbf{16} & = (\textbf{3} ,{+\textstyle\frac{1}{2}}, 
\textbf{2} , \textbf{1}) +
(\overline{\textbf{3}} ,{-\textstyle\frac{1}{2}}, \textbf{1} , 
\textbf{2}) + (\textbf{1}
,{-\textstyle\frac{3}{2}}, \textbf{2} , \textbf{1}) + 
(\textbf{1} ,{+\textstyle\frac{3}{2}}, \textbf{1} ,
\textbf{2}) \\
& = Q_L + Q_R + L_L + L_R \\
 \overline{\textbf{16}} & = (\overline{\textbf{3}} ,
{-\textstyle\frac{1}{2}}, \textbf{2} ,
\textbf{1}) + (\textbf{3} ,{+\textstyle\frac{1}{2}}, \textbf{1} ,
\textbf{2}) + (\textbf{1} ,{+\textstyle\frac{3}{2}}, \textbf{2} ,
\textbf{1}) + (\textbf{1} ,{-\textstyle\frac{3}{2}}, \textbf{1} ,
\textbf{2}) \\
& = \overline{Q}_L + \overline{Q}_R + \overline{L}_L + \overline{L}_R
\end{split}
\end{equation}
A model must then have three families in order to be phenomenologically viable
\textit{i.e}
\begin{equation}
N_{Q_L} - N_{\overline{Q}_L} = N_{Q_R} - N_{\overline{Q}_R} = N_{L_L} -
N_{\overline{L}_L} = N_{L_R} - N_{\overline{L}_R} = 3
\end{equation}
The number of these representations that occur in a model depends on the choice
of the GGSO coefficients. Firstly, in order to distinguish between the
\textbf{16} and $\overline{\textbf{16}}$ an $SO(10)$ chirality operator is
defined. These chirality operators for the sectors $B^{(1,2,3)}_{pqrs}$ are
defined, respectively, as
\begin{equation}
\begin{split}
& X^{(1)SO(10)}_{pqrs} = C\binom{B^{(1)}_{pqrs}}{(1-r)e_5 + (1-s)e_6 + b_2} \\
& X^{(2)SO(10)}_{pqrs} = C\binom{B^{(2)}_{pqrs}}{(1-r)e_5 + (1-s)e_6 + b_1} \\
& X^{(3)SO(10)}_{pqrs} = C\binom{B^{(3)}_{pqrs}}{(1-r)e_3 + (1-s)e_4 + b_1}
\end{split}
\end{equation}
and can take the values $X^{(1,2,3)SO(10)}_{pqrs} = \pm 1$. Another chirality
operator needs defining to determine whether the representations
$((\textbf{1},\textbf{2})$ or $(\textbf{2},\textbf{1}))$ of the $SU(2)_L\times
SU(2)_R$ occur. These are defined for the sectors $B^{(1,2,3)}_{pqrs}$ 
respectively as
\begin{equation}
\begin{split}
& X^{(1) SU(2)_{L/R}}_{pqrs} = C\binom{B^{(1)}_{pqrs}}{2\alpha + x} \\
& X^{(2) SU(2)_{L/R}}_{pqrs} = C\binom{B^{(2)}_{pqrs}}{2\alpha + x} \\
& X^{(3) SU(2)_{L/R}}_{pqrs} = C\binom{B^{(3)}_{pqrs}}{2\alpha + x}
\end{split}
\end{equation}
where $x$ is the linear combination $x = \mathds{1} + S + \sum^{6}_{i=1} e_i +
z_1 + z_2$.

Furthermore, there is one final chirality operator which needs to be defined in
order to determine the repesentations under the $SU(3)_C\times U(1)_C$ gauge
group. These are
\begin{equation}
\begin{split}
& X^{(1) SU(3)\times U(1)}_{pqrs} = C\binom{B^{(1)}_{pqrs}}{(1-p)e_3 + 
(1-q)e_4 + b_3 + x + 2\alpha} \\
& X^{(2) SU(3)\times U(1)}_{pqrs} = C\binom{B^{(2)}_{pqrs}}{(1-p)e_1 + 
(1-q)e_2 + b_3 + x + 2\alpha} \\
& X^{(3) SU(3)\times U(1)}_{pqrs} = C\binom{B^{(3)}_{pqrs}}{(1-p)e_1 + 
(1-q)e_2 + b_2 + x + 2\alpha}
\end{split}
\end{equation}
By performing the GSO projections of these chirality operators the
surviving states and therefore the number of families are calculated.

\subsubsection{Projectors}
The projectors are a set of equations which determine whether a sector is
either projected out or kept in the string spectrum. These projectors consist
of the relevant GGSO coefficients for the sector.
For the observable chiral matter there are 48 projectors which are calculated
to be
\footnotesize
\begin{equation}
\begin{split}
& P^{(1)}_{pqrs} = \frac{1}{16}\biggl(1 -
C\binom{e_1}{B^{(1)}_{pqrs}}\biggr)\cdot\biggl(1 -
C\binom{e_2}{B^{(1)}_{pqrs}}\biggr)\cdot\biggl(1 -
C\binom{z_1}{B^{(1)}_{pqrs}}\biggr)\cdot\biggl(1 -
C\binom{z_2}{B^{(1)}_{pqrs}}\biggr) \\
& P^{(2)}_{pqrs} = \frac{1}{16}\biggl(1 -
C\binom{e_3}{B^{(2)}_{pqrs}}\biggr)\cdot\biggl(1 -
C\binom{e_4}{B^{(2)}_{pqrs}}\biggr)\cdot\biggl(1 -
C\binom{z_1}{B^{(2)}_{pqrs}}\biggr)\cdot\biggl(1 -
C\binom{z_2}{B^{(2)}_{pqrs}}\biggr) \\
& P^{(3)}_{pqrs} = \frac{1}{16}\biggl(1 -
C\binom{e_5}{B^{(3)}_{pqrs}}\biggr)\cdot\biggl(1 -
C\binom{e_6}{B^{(3)}_{pqrs}}\biggr)\cdot\biggl(1 -
C\binom{z_1}{B^{(3)}_{pqrs}}\biggr)\cdot\biggl(1 -
C\binom{z_2}{B^{(3)}_{pqrs}}\biggr)
\end{split}
\end{equation}
\normalsize
The analysis of the physical spectrum is formulated as algebraic equations. 
The projectors can be expressed
as a system of linear equations where $p,q,r,s$ take unknown values. 
The sectors which survive the GSO projections are found by
solving the systems of equations for $p,q,r,s$. 
Using this formalism allows for a
computer analysis of the models as the systems of linear equations
are easy to express in a computer code.

The following notation is used in the algebraic representation
of the GGSO projections
\begin{equation}
C\binom{v_i}{v_j} = e^{i\pi(v_i | v_j)}.
\end{equation}
When the GGSO coefficients are expressed in this way the analytic expressions
for the projectors $P^{(1,2,3)}_{pqrs}$ are given in matrix form $\Delta^i W^i
= Y^i$ as
\begin{equation*}
\begin{pmatrix}
(e_1|e_3) & (e_1|e_4) & (e_1|e_5) & (e_1|e_6)\\
(e_2|e_3) & (e_2|e_4) & (e_2|e_5) & (e_2|e_6)\\
(z_1|e_3) & (z_1|e_4) & (z_1|e_5) & (z_1|e_6)\\
(z_2|e_3) & (z_2|e_4) & (z_2|e_5) & (z_2|e_6)\\
\end{pmatrix}
\begin{pmatrix}
p\\ q \\ r \\s
\end{pmatrix}
=
\begin{pmatrix}
(e_1|b_1)\\
(e_2|b_1)\\
(z_1|b_1)\\
(z_2|b_1)
\end{pmatrix}
\end{equation*}
\begin{equation*}
\begin{pmatrix}
(e_3|e_1) & (e_3|e_2) & (e_3|e_5) & (e_3|e_6)\\
(e_4|e_1) & (e_4|e_2) & (e_4|e_5) & (e_4|e_6)\\
(z_1|e_1) & (z_1|e_2) & (z_1|e_5) & (z_1|e_6)\\
(z_2|e_1) & (z_2|e_2) & (z_2|e_5) & (z_2|e_6)
\end{pmatrix}
\begin{pmatrix}
p\\ q \\ r \\s
\end{pmatrix}
=
\begin{pmatrix}
(e_3|b_2)\\
(e_4|b_2)\\
(z_1|b_2)\\
(z_2|b_2)
\end{pmatrix}
\end{equation*}
\begin{equation*}
\begin{pmatrix}
(e_5|e_1) & (e_5|e_2) & (e_5|e_3) & (e_5|e_4)\\
(e_6|e_1) & (e_6|e_2) & (e_6|e_3) & (e_6|e_4)\\
(z_1|e_1) & (z_1|e_2) & (z_1|e_3) & (z_1|e_4)\\
(z_2|e_1) & (z_2|e_2) & (z_2|e_3) & (z_2|e_4)
\end{pmatrix}
\begin{pmatrix}
p\\ q \\ r \\s
\end{pmatrix}
=
\begin{pmatrix}
(e_5|b_3)\\
(e_6|b_3)\\
(z_1|b_3)\\
(z_2|b_3)
\end{pmatrix}
\end{equation*}
respectively. Such algebraic matrix equations can be written 
for the entire physical spectrum. In the ensuing discussion we
list all the sectors that can a priori produce physical states, 
but do not list explicitly all the algebraic 
matrix equations for the corresponding GGSO projections.

\subsection{Exotic Sectors}
Additional sectors exist in the string models that can give rise to states
that carry fractional charges under the LRS gauge group. 
This leads to states
with a fractional electric charge at the level of the Standard Model. The term
`exotic states' used here is reserved purely for the states with fractional
electric charge which arise from the sectors containing the basis vector
$\alpha$. Exotic states arise from these sectors due to Wilson line breaking
of the non-Abelian GUT symmetries. These exotics states are a generic feature
of string compactifications \cite{25, 26, 27} 
and experimental searches are being
conducted in order to find them \cite{28}. There are interesting
phenomenological aspects to exotic states as charge conservation implies that
the lightest of these states is necessarily stable. To date however, no such
exotic states have been observed, leading to strong upper bounds on their
abundance \cite{28}. In addition, if these states are too plentiful in the
early universe they can cause problems during the reheating phase as the
lightest of these states is necessarily stable, meaning they continue to
scatter and cannot decouple from the plasma in the early Universe due to their
charge.

There are two solutions to the lack of experimental data for the existence of
exotics. The first solution is by demanding that the exotics are confined to
integrally charged states \cite{fsu5}. The second is to demand that the exotic
states are sufficiently heavy and diluted in the cosmological evolution of the
universe \cite{27}. However, there are issues with the integrally charged state
solution as these states affect the renormalisation group running of the
weak-hypercharge and gauge group unification. This leads to the preferred
solution of demanding that 
the exotic states are sufficiently massive and dilute. A
sufficient mass for these states is above the GUT scale so that they are
diluted during the inflationary period of the universe as during the reheating
phase they will not be reproduced.

Previous classifications of heterotic-string models found 
examples of vacua in which massless exotics states
were absent and only appeared in the massive spectrum.  
These models were dubbed `exophobic heterotic string vacua'. 
In the case of the Pati-Salam models, three generation
exophobic vacua were found \cite{acfkr} and in the FSU5 case exophobic
vacua were found in models with an even number of generations \cite{frs}. A
question of interest for the current research is therefore whether any
exophobic LRS models can be found.

\subsubsection{Spinorial Exotics}\label{spinorialexotics}
The term spinorial exotics refers to sectors which involve the basis vector
$\alpha$ and have the products $\xi_L \cdot \xi_L = 4$ and 
$\xi_R \cdot \xi_R =8$,
therefore requiring no oscillators to produce massless states.

The sectors below all give rise to states with the representations
$(\textbf{1},-\frac{3}{4},\textbf{1},\textbf{2})$ and
$(\textbf{1},-\frac{3}{4},\textbf{2},\textbf{1})$ under the $SU(3)_C \times
U(1)_C \times SU(2)_L \times SU(2)_R$ observable gauge group. These states are
defined in the analysis as $n_{L_Le}$ and $n_{L_Re}$ respectively. It can be
seen that these are singlets under the $SU(3)_C$ gauge group but are still
charged under $U(1)_C$. The corresponding sectors with $3\alpha$
in the linear combination of basis vectors give states with the
representations $(\textbf{1},+\frac{3}{4},\textbf{1},\textbf{2})$ and
$(\textbf{1},+\frac{3}{4},\textbf{2},\textbf{1})$. It can be seen that the only
change is the sign reversal of the charge under $U(1)_C$. The following are
the sectors which give rise to these representations
\begin{eqnarray}
B_{pqrs}^{(7)} &=& B_{pqrs}^{(1)} + \alpha
\nonumber \\ 
&=& \{\psi^{\mu},\chi^{1,2},(1-p)y^3\bar{y}^3,pw^3\bar{w}^3,(1-q)y^4\bar{y}^4,qw^4\bar{w}^4,
\nonumber \\
& & ~~~ (1-r)y^5\bar{y}^5,rw^5\bar{w}^5,(1-s)y^6\bar{y}^6,sw^6\bar{w}^6,\bar{\eta}^1 = -\textstyle\frac{1}{2},
\\
& & ~~~~~~~~~~~~~~~~ \bar{\eta}^{2,3} = \textstyle\frac{1}{2},\bar{\psi}^{1,2,3} = -\textstyle\frac{1}{2},\bar{\psi}^{4,5},\bar{\phi}^{1,\ldots,6} = \textstyle\frac{1}{2},\bar{\phi}^7\}
\nonumber \\
B_{pqrs}^{(8,9)} &=& B^{(2,3)}_{pqrs} + \alpha \nonumber
\end{eqnarray}
\begin{eqnarray}
B_{pqrs}^{(13)} &=& B_{pqrs}^{(1)} + z_1 + \alpha
\nonumber \\ 
&=& \{\psi^{\mu},\chi^{1,2},(1-p)y^3\bar{y}^3,pw^3\bar{w}^3,(1-q)y^4\bar{y}^4,qw^4\bar{w}^4,
\nonumber \\
& & ~~~ (1-r)y^5\bar{y}^5,rw^5\bar{w}^5,(1-s)y^6\bar{y}^6,sw^6\bar{w}^6,\bar{\eta}^1 = -\textstyle\frac{1}{2},
\\
& & ~~~~~~ \bar{\eta}^{2,3} = \textstyle\frac{1}{2},\bar{\psi}^{1,2,3} = -\textstyle\frac{1}{2},\bar{\psi}^{4,5},\bar{\phi}^{1,\ldots,4} = -\textstyle\frac{1}{2},\bar{\phi}^{5,6} = \textstyle\frac{1}{2},\bar{\phi}^7\}
\nonumber \\
B_{pqrs}^{(14,15)} &=& B^{(2,3)}_{pqrs} + z_1 + \alpha \nonumber
\end{eqnarray}
\begin{eqnarray}
B_{pqrs}^{(22)} &=& B_{pqrs}^{(1)} + z_2 + \alpha
\nonumber \\ 
&=& \{\psi^{\mu},\chi^{1,2},(1-p)y^3\bar{y}^3,pw^3\bar{w}^3,(1-q)y^4\bar{y}^4,qw^4\bar{w}^4,
\nonumber \\
& & ~~~ (1-r)y^5\bar{y}^5,rw^5\bar{w}^5,(1-s)y^6\bar{y}^6,sw^6\bar{w}^6,\bar{\eta}^1 = -\textstyle\frac{1}{2},
\\
& & ~~~~~~ \bar{\eta}^{2,3} = \textstyle\frac{1}{2},\bar{\psi}^{1,2,3} = -\textstyle\frac{1}{2},\bar{\psi}^{4,5},\bar{\phi}^{1,\ldots,4} = \textstyle\frac{1}{2},\bar{\phi}^{5,6} = -\textstyle\frac{1}{2},\bar{\phi}^8\}
\nonumber \\
B_{pqrs}^{(23,24)} &=& B^{(2,3)}_{pqrs} + z_2 + \alpha \nonumber
\end{eqnarray}
\begin{eqnarray}
B_{pqrs}^{(31)} &=& B_{pqrs}^{(1)} + z_1 + z_2 + \alpha
\nonumber \\ 
&=& \{\psi^{\mu},\chi^{1,2},(1-p)y^3\bar{y}^3,pw^3\bar{w}^3,(1-q)y^4\bar{y}^4,qw^4\bar{w}^4,
\nonumber \\
& & ~~~ (1-r)y^5\bar{y}^5,rw^5\bar{w}^5,(1-s)y^6\bar{y}^6,sw^6\bar{w}^6,\bar{\eta}^1 = -\textstyle\frac{1}{2},
\\
& & ~~~~~~ \bar{\eta}^{2,3} = \textstyle\frac{1}{2},\bar{\psi}^{1,2,3} = -\textstyle\frac{1}{2},\bar{\psi}^{4,5},\bar{\phi}^{1,\ldots,4} = -\textstyle\frac{1}{2},\bar{\phi}^{5,6} = -\textstyle\frac{1}{2},\bar{\phi}^8\}
\nonumber \\
B_{pqrs}^{(32,33)} &=& B^{(2,3)}_{pqrs} + z_1 + z_2 + \alpha \nonumber
\end{eqnarray}

\subsubsection{Vectorial Exotics}\label{vectorialexotics}
The following are vectorial states, meaning they have the products $\xi_L \cdot
\xi_L = 4$ and $\xi_R \cdot \xi_R = 6$, therefore requiring 
one $\frac{1}{4}$
oscillator to produce massless states. Firstly, there are the sectors
\begin{eqnarray}
B_{pqrs}^{(46)} &=& B_{pqrs}^{(1)} + \alpha + x
\nonumber \\ 
&=& \{\psi^{\mu},\chi^{1,2},(1-p)y^3\bar{y}^3,pw^3\bar{w}^3,
(1-q)y^4\bar{y}^4,qw^4\bar{w}^4,
\nonumber \\
& & ~~~ (1-r)y^5\bar{y}^5,rw^5\bar{w}^5,(1-s)y^6\bar{y}^6,
sw^6\bar{w}^6,\bar{\eta}^1 = \textstyle\frac{1}{2},
\\
& & ~~~~~~~~~~~~~~~~~~ \bar{\eta}^{2,3} = -\textstyle\frac{1}{2},
\bar{\psi}^{1,2,3} = \textstyle\frac{1}{2},\bar{\phi}^{1,\ldots,6} = 
\textstyle\frac{1}{2},\bar{\phi}^7\}
\nonumber \\
B_{pqrs}^{(47,48)} &=& B^{(2,3)}_{pqrs} + \alpha + x \nonumber
\end{eqnarray}

Using $B^{(46)}_{pqrs}$ as an example to show the states that 
can be obtained from
these sectors, the possible states are
\begin{itemize}
\item $\{\overline{\psi}^{*1,2,3} \} \ket{R}^{(46)}_{pqrs}$, where
$\ket{R}^{(46)}_{pqrs}$ is the degenerate Ramond vacua of the $B^{(46)}_{pqrs}$
sector. These states transform as vector-like representations of the observable
$SU(3)_C \times U(1)_C$.
\item $\{\overline{\eta}^{*1}\} \ket{R}^{(46)}_{pqrs}$. 
These states transform as vector-like representations
of $U(1)_1$.
\item $\{\overline{\eta}^{2,3}\} \ket{R}^{(46)}_{pqrs}$. These states transform
as vector-like representations of $U(1)_2$ and $U(1)_3$ respectively.
\item $\{\overline{\phi}^{*1,\ldots,4} \}\ket{R}^{(46)}_{pqrs}$. These states
transform as vector-like representations of the hidden
 $SU(4)\times U(1)_4$.
\item $\{\overline{\phi}^{*5,6} \}\ket{R}^{(46)}_{pqrs}$. These states
transform as vector-like representations of the hidden
$SU(2)_5\times U(1)_5$.
\end{itemize}
The states obtained from the sectors
$B^{(47,48)}_{pqrs}$ transform in the same manner as those above.

Secondly, there are the following 48 sectors
\begin{eqnarray}
B_{pqrs}^{(52)} &=& B_{pqrs}^{(1)} + z_1 + \alpha + x
\nonumber \\ 
&=& \{\psi^{\mu},\chi^{1,2},(1-p)y^3\bar{y}^3,pw^3\bar{w}^3,
(1-q)y^4\bar{y}^4,qw^4\bar{w}^4,
\nonumber \\
& & ~~~ (1-r)y^5\bar{y}^5,rw^5\bar{w}^5,(1-s)y^6\bar{y}^6,
sw^6\bar{w}^6,\bar{\eta}^1 = \textstyle\frac{1}{2},
\\
& & ~~~~~ \bar{\eta}^{2,3} = -\textstyle\frac{1}{2},
\bar{\psi}^{1,2,3} = \textstyle\frac{1}{2},\bar{\phi}^{1,\ldots,4} = 
-\textstyle\frac{1}{2},\bar{\phi}^{5,6} = \textstyle\frac{1}{2},\bar{\phi}^7\}
\nonumber \\
B_{pqrs}^{(53,54)} &=& B^{(2,3)}_{pqrs} + z_1 + \alpha + x \nonumber
\end{eqnarray}
The states found from these sectors only differ from $B^{(47,48,49)}_{pqrs}$ by
a negative sign on the $\frac{1}{2}$ boundary conditions of the fermions
$\overline{\phi}^{1,2,3,4}$. This has the effect of changing the sign of the
$U(1)_4$ charges while leaving the other charges unaffected. The structure and
charges generated by the other worldsheet
fermions therefore remain identical.

Similarly, in the sectors
\begin{eqnarray}
B_{pqrs}^{(58)} &=& B_{pqrs}^{(1)} + z_2 + \alpha + x
\nonumber \\ 
&=& \{\psi^{\mu},\chi^{1,2},(1-p)y^3\bar{y}^3,pw^3\bar{w}^3,
(1-q)y^4\bar{y}^4,qw^4\bar{w}^4,
\nonumber \\
& & ~~~ (1-r)y^5\bar{y}^5,rw^5\bar{w}^5,(1-s)y^6\bar{y}^6,
sw^6\bar{w}^6,\bar{\eta}^1 = \textstyle\frac{1}{2},
\\
& & ~~~~~ \bar{\eta}^{2,3} = -\textstyle\frac{1}{2},
\bar{\psi}^{1,2,3} = \textstyle\frac{1}{2},\bar{\phi}^{1,\ldots,4} = 
\textstyle\frac{1}{2},\bar{\phi}^{5,6} = 
-\textstyle\frac{1}{2},\bar{\phi}^8\}
\nonumber \\
B_{pqrs}^{(59,60)} &=& B^{(2,3)}_{pqrs} + z_2 + \alpha + x, \nonumber
\end{eqnarray}
the observable states are identical to those in the sectors
$B^{(47,48,49)}_{pqrs}$ and only
the hidden charges differ by a slight change in the Ramond vacua and a sign
difference of the boundary conditions of the fermions $\overline{\phi}^{5,6}$,
which only affects the sign of the charges under $U(1)_5$.

The final 48 sectors are
\begin{eqnarray}
B_{pqrs}^{(64)} &=& B_{pqrs}^{(1)} + z_1 + z_2 + \alpha + x
\nonumber \\ 
&=& \{\psi^{\mu},\chi^{1,2},(1-p)y^3\bar{y}^3,pw^3\bar{w}^3,
(1-q)y^4\bar{y}^4,qw^4\bar{w}^4,
\nonumber \\
& & ~~~ (1-r)y^5\bar{y}^5,rw^5\bar{w}^5,(1-s)y^6\bar{y}^6,
sw^6\bar{w}^6,\bar{\eta}^1 = \textstyle\frac{1}{2},
\\
& & ~~~~~~~~~~~~~~~~~~ \bar{\eta}^{2,3} = 
-\textstyle\frac{1}{2},\bar{\psi}^{1,2,3} = 
\textstyle\frac{1}{2},\bar{\phi}^{1,\ldots,6} = 
-\textstyle\frac{1}{2},\bar{\phi}^8\}
\nonumber \\
B_{pqrs}^{(65,66)} &=& B^{(2,3)}_{pqrs} + z_1 + z_2 + \alpha + x \nonumber
\end{eqnarray}
These differ from sectors $B^{(58,59,60)}_{pqrs}$ by changing the sign on the
$\frac{1}{2}$ boundary conditions of the fermions $\overline{\phi}^{1,2,3,4}$
and therefore, as above, there is a sign change on the charges under $U(1)_4$.
All other states are unaffected and remain as in the sectors
$B^{(58,59,60)}_{pqrs}$.

\subsubsection{Pati-Salam Exotics}\label{patisalamexotics}
In the case of left-right symmetric models, there can be states which are
exotic with respect to the Pati-Salam gauge group $SO(6)\times SO(4)$. The
sectors from which these states can arise are those which contain the
vector combination 
$2\alpha$. This is due to the fermions $\overline{\psi}^{1,2,3}$ or
$\overline{\psi}^{4,5}$ having periodic boundary conditions in the sector
(therefore generating the Pati-Salam gauge subgroup), while still having a
fractional electric charge with respect to the Standard Model.

In the model being discussed, all of the Pati-Salam exotics are found in the
following sectors:
\begin{eqnarray}
B_{pqrs}^{(70)} &=& B_{pqrs}^{(1)} + z_1 + 2\alpha
\nonumber \\ 
&=& \{\psi^{\mu},\chi^{1,2},(1-p)y^3\bar{y}^3,pw^3\bar{w}^3,
(1-q)y^4\bar{y}^4,qw^4\bar{w}^4,
 \\
& & ~~~ (1-r)y^5\bar{y}^5,rw^5\bar{w}^5,(1-s)y^6\bar{y}^6,
sw^6\bar{w}^6,\bar{\eta}^{2,3},\bar{\psi}^{4,5},\bar{\phi}^{5,6}\}
\nonumber \\
B_{pqrs}^{(71,72)} &=& B^{(2,3)}_{pqrs} + z_1 + 2\alpha \nonumber
\end{eqnarray}
These states transform in representations of the gauge group $SU(2)_L \times
SU(2)_R\times SU(2)_5 \times U(1)_5$.
\begin{eqnarray}
B_{pqrs}^{(34)} &=& B_{pqrs}^{(1)} + z_1 + z_2 + 2\alpha
\nonumber \\ 
&=& \{\psi^{\mu},\chi^{1,2},(1-p)y^3\bar{y}^3,pw^3\bar{w}^3,
(1-q)y^4\bar{y}^4,qw^4\bar{w}^4,
 \\
& & ~~~ (1-r)y^5\bar{y}^5,rw^5\bar{w}^5,(1-s)y^6\bar{y}^6,
sw^6\bar{w}^6,\bar{\eta}^{2,3},\bar{\psi}^{4,5},\bar{\phi}^{7,8}\}
\nonumber \\
B_{pqrs}^{(35,36)} &=& B^{(2,3)}_{pqrs} + z_1 + z_2+ 2\alpha \nonumber
\end{eqnarray}
These states transform as representations of the gauge group $SU(2)_L \times
SU(2)_R \times U(1)_7 \times U(1)_8$. The states from the previous 96 sectors
are defined in the analysis as $n_{L_Ls} , n_{L_Rs} , n_{\overline{L}_Ls}$ and
$n_{\overline{L}_Rs}$.
\begin{eqnarray}
B_{pqrs}^{(40)} &=& B_{pqrs}^{(1)} + z_1 + 2\alpha + x
\nonumber \\ 
&=& \{\psi^{\mu},\chi^{1,2},(1-p)y^3\bar{y}^3,pw^3\bar{w}^3,
(1-q)y^4\bar{y}^4,qw^4\bar{w}^4,
 \\
& & ~~~ (1-r)y^5\bar{y}^5,rw^5\bar{w}^5,(1-s)y^6\bar{y}^6,
sw^6\bar{w}^6,\bar{\eta}^{1},\bar{\psi}^{1,2,3},\bar{\phi}^{5,6}\}
\nonumber \\
B_{pqrs}^{(41,42)} &=& B^{(2,3)}_{pqrs} + z_1 + 2\alpha + x \nonumber
\end{eqnarray}
These states transform as representations of the gauge group 
$SU(3)_C \times U(1)_C \times SU(2)_5 \times U(1)_5$
\begin{eqnarray}
B_{pqrs}^{(43)} &=& B_{pqrs}^{(1)} + z_1 + z_2 + 2\alpha + x
\nonumber \\ 
&=& \{\psi^{\mu},\chi^{1,2},(1-p)y^3\bar{y}^3,pw^3\bar{w}^3,
(1-q)y^4\bar{y}^4,qw^4\bar{w}^4,
 \\
& & ~~~ (1-r)y^5\bar{y}^5,rw^5\bar{w}^5,(1-s)y^6\bar{y}^6,
sw^6\bar{w}^6,\bar{\eta}^{1},\bar{\psi}^{1,2,3},\bar{\phi}^{7,8}\}
\nonumber \\
B_{pqrs}^{(44,45)} &=& B^{(2,3)}_{pqrs} + z_1 + z_2 + 2\alpha + x \nonumber
\end{eqnarray}
These states transform as representations of the gauge group $SU(3)_C \times
U(1)_C \times U(1)_7 \times U(1)_8$. The states from the previous 96 sectors
are defined in the analysis as $n_{3v}$ and $n_{\overline{3}v}$.

\subsection{Hidden Matter Spectrum}\label{hiddenmatter}
The hidden matter spectrum refers to sectors which produce states that
transform under the hidden gauge group but are singlets under the observable
SO(10) GUT gauge group. This means that the states produced are not exotic with
respect to the Standard Model gauge charges.

There are 48 sectors present from $B^{(1,2,3)}_{pqrs} + z_1 + x$ which are
\begin{eqnarray}
B_{pqrs}^{(19)} &=& B_{pqrs}^{(1)} + z_1 + x
\nonumber \\ 
&=& \{\psi^{\mu},\chi^{1,2},(1-p)y^3\bar{y}^3,pw^3\bar{w}^3,
(1-q)y^4\bar{y}^4,qw^4\bar{w}^4,
 \\
& & ~~~ (1-r)y^5\bar{y}^5,rw^5\bar{w}^5,(1-s)y^6\bar{y}^6,
sw^6\bar{w}^6,\bar{\eta}^{2,3},\bar{\phi}^{1,2,3,4}\}
\nonumber \\
B_{pqrs}^{(20,21)} &=& B^{(2,3)}_{pqrs} + z_1 + x \nonumber
\end{eqnarray}
These sectors contain states which transform under the hidden
$SU(4) \times U(1)_4$
gauge group with the representations $(\textbf{1},+2)$, $(\textbf{1},-2)$,
$(\textbf{4},+1)$, $(\overline{\textbf{4}},-1)$, $(\textbf{6},0)$.

There exists another 48 sectors $B^{(1,2,3)}_{pqrs} + z_2 + x$ given by
\begin{eqnarray}
B_{pqrs}^{(28)} &=& B_{pqrs}^{(1)} + z_2 + x
\nonumber \\ 
&=& \{\psi^{\mu},\chi^{1,2},(1-p)y^3\bar{y}^3,pw^3\bar{w}^3,
(1-q)y^4\bar{y}^4,qw^4\bar{w}^4,
 \\
& & ~~~ (1-r)y^5\bar{y}^5,rw^5\bar{w}^5,(1-s)y^6\bar{y}^6,
sw^6\bar{w}^6,\bar{\eta}^{2,3},\bar{\phi}^{5,6,7,8}\}
\nonumber \\
B_{pqrs}^{(29,30)} &=& B^{(2,3)}_{pqrs} + z_2 + x \nonumber
\end{eqnarray}
These sectors produce states which transform under the $SU(2)_5 \times U(1)_5
\times U(1)_7 \times U(1)_8$ gauge group with the representations:
$(\textbf{1}, +1, \pm\frac{1}{2}, \pm\frac{1}{2})$ , $(\textbf{2}, 0,
\pm\frac{1}{2}, \pm\frac{1}{2})$ , $(\textbf{1}, -1, \pm\frac{1}{2},
\pm\frac{1}{2})$ where the charges of $U(1)_7$ and $U(1)_8$ can take all
possible permutations of the values given, meaning there are 12 distinct
representations in total.

\section{Classification Results and Analysis}\label{CRAL}
The classification process involves utilising the calculated algebraic
conditions which were presented in the previous sections. By using the
projectors and chirality operators for each sector the entire massless spectrum
can be analysed for a specific choice of the one-loop GGSO projection
coefficients. These algebraic conditions can be written in a computer program
which enables a scan over the different choices of GGSO projection
coefficients. As the total number of possible configurations, and therefore
vacua, is $2^{66} \approx 7.38\times 10^{19}$ a complete scan of the entire
space of 
string vacua is not possible. Therefore, a random generation of the GGSO
projection coefficients is used in order to provide a random sample of
vacua\footnote[3]{We note here that analysis of large sets of string vacua
have been performed by other research groups \cite{statistical}}
from which models with desirable phenomenological criteria can be found.

The algebraic conditions were programmed into a JAVA code in order to perform
the classification and the accuracy of this program was checked against an
independently written FORTRAN code. In the JAVA program, a random generator was
used in order to provide the different GGSO configurations. This
program initially produces a random GGSO configuration, before running these
values through the algebraic conditions calculated for each sector in order to
produce the full spectrum of each model. By repeating this process, the
statistics associated with classification can be developed while also fishing
for single models which are of phenomenological significance.

Previous papers which have utilised this technique can be seen in
references \cite{acfkr, 421, frs, SLM}. In the case of the classification
of Pati-Salam models, this method was shown to produce three-generation models
which contained no exotic massless states with fractional electric charge, and
were therefore exophobic.

Therefore, an example of a question of phenomenological interest is whether
exophobic LRS models can be found.

\begin{table}
\begin{tabular}{cccc}
\toprule
Spinorial $SO(10)$ & Vectorial $SO(10)$ & LRS Exotic & Pati-Salam Exotic \\
Observable & Observable\\
\midrule
$n_{L_L} = (\textbf{1}, -\nicefrac{3}{2} , \textbf{2} , \textbf{1} )$ & $n_{h} = (\textbf{1} , 0 ,
\textbf{2} , \textbf{2} )$ & $n_{L_L s} = ( \textbf{1}, +\nicefrac{3}{4} , \textbf{2} ,
\textbf{1} )$ & $n_{L_L e} = (\textbf{1}, 0 , \textbf{2} , \textbf{1} )$ \\
$n_{L_R} = (\textbf{1}, +\nicefrac{3}{2} , \textbf{1} , \textbf{2} )$ & $n_{3} = (\textbf{3}, -1 ,
\textbf{1} , \textbf{1} )$ & $n_{L_R s} = ( \textbf{1}, +\nicefrac{3}{4} , \textbf{1} ,
\textbf{2} )$& $n_{L_R e} = ( \textbf{1}, 0 , \textbf{1} , \textbf{2} )$\\
$n_{Q_L} = (\textbf{3}, +\nicefrac{1}{2} , \textbf{2} , \textbf{1} )$ & $n_{\overline{3}} =
(\overline{\textbf{3}}, +1 , \textbf{1} , \textbf{1} )$& $n_{\overline{L}_L s} = (
\textbf{1}, -\nicefrac{3}{4} , \textbf{2} , \textbf{1} )$& $n_{3e} = ( \textbf{3}, +\nicefrac{1}{2} , \textbf{1} ,
\textbf{1} )$\\
$n_{Q_R} = (\overline{\textbf{3}}, -\nicefrac{1}{2} , \textbf{1} , \textbf{2} )$& &
$n_{\overline{L}_R s} = ( \textbf{1}, -\nicefrac{3}{4} , \textbf{1} , \textbf{2} )$&
$n_{\overline{3}e} = ( \overline{\textbf{3}}, +\nicefrac{1}{2} , \textbf{1} , \textbf{1} )$\\
$n_{\overline{L}_L} = (\textbf{1}, +\nicefrac{3}{2} , \textbf{2} , \textbf{1} )$& &$n_{3v} = (
\textbf{3}, +\nicefrac{1}{4} , \textbf{1} , \textbf{1} )$ & $n_{1e} = ( \textbf{1}, +\nicefrac{3}{2} , \textbf{1} ,
\textbf{1} )$\\
$n_{\overline{L}_R} = (\textbf{1}, -\nicefrac{3}{2} , \textbf{1} , \textbf{2} )$& &
$n_{\overline{3}v} = ( \overline{\textbf{3}}, -\nicefrac{1}{4} , \textbf{1} , \textbf{1} )$ &
$n_{\overline{1}e} = ( \textbf{1}, -\nicefrac{3}{2} , \textbf{1} , \textbf{1} )$\\
$n_{\overline{Q}_L} = (\overline{\textbf{3}}, -\nicefrac{1}{2} , \textbf{2} , \textbf{1} )$ & & $n_{1 v} = (\textbf{1}, +\nicefrac{3}{4}, \textbf{1}, \textbf{1})$\\
$n_{\overline{Q}_R} = (\textbf{3}, +\nicefrac{1}{2} , \textbf{1} , \textbf{2} )$ & & $n_{\overline{1} v} = (\textbf{1}, -\nicefrac{3}{4}, \textbf{1}, \textbf{1})$\\
\midrule 
\multicolumn{4}{c}{$n_g = n_{L_L} - n_{\overline{L}_L} = n_{L_R} - n_{\overline{L}_R} = n_{Q_L} -
n_{\overline{Q}_L} = n_{Q_R} - n_{\overline{Q}_R}$} \\
\multicolumn{2}{c}{$n_{H} = n_{\overline{L}_R}$~~~~~}\\
\bottomrule
\end{tabular}
\caption{\label{phenonumbers} \emph{The 27 integers used to catergorise the
quantities of phenomenological interest. The first column contains states from
the \textbf{16} and $\overline{\textbf{16}}$ representations of $SO(10)$. The second contains
the states from the $\textbf{10}$ representation of $SO(10)$. The third and fourth list the
states which are exotic with respect to the Left-Right Symmetric and Pati-Salam
gauge groups respectively.}}
\end{table}

The observable sector of a heterotic-string Left-Right Symmetric model is
characterised by 27 integers which are defined in table \ref{phenonumbers}.
These contain the relevent quantities of phenomenological interest.
Notable numbers defined in table \ref{phenonumbers} are $n_g$, $n_h$
and $n_H$ as these give the number of generations of a model and
whether the model contains non-chiral light and heavy Higgs representations.

The numbers given in the first two columns of table \ref{phenonumbers}
are as described above in section
(4.2). The first four numbers form a complete \textbf{16} of $SO(10)$ and the
last four form a complete $\overline{\textbf{16}}$. The first four in the LRS
Exotics column arise from the spinorial exotic sectors and the last two arise
from the vectorial exotic sectors. 
To perform the classification, the analytic formulae for all the sectors
which contribute to these numbers were derived so as to describe the complete
spectrum of each model.

For a model to be phenomenologically viable, it must satisfy the following
phenomenological criteria:
\begin{eqnarray*}
& n_g = 3 &  \text{Three light chiral generations} \\
& n_{H} \geq 1 & \text{At least one heavy Higgs pair to break the $SU(2)_R$
symmetry}\\
& n_h \geq 1 &  \text{At least one light Higgs bi-doublet} \\
& n_3 = n_{\overline{3}} &  \text{Heavy mass can be generated for the
colour triplets}\\
& n_{3e} = n_{\overline{3}e} &  \text{Heavy mass can be generated for
the colour triplets}\\
& n_{1e} = n_{\overline{1}e} & \text{Heavy mass can be generated for vector-like
exotics}\\
& n_{3v} = n_{\overline{3}v} &  \text{Heavy mass can be generated for
the colour triplets}\\
& n_{1v} = n_{\overline{1}v} &  \text{Heavy mass can be generated for
the vector-like exotics}\\
& n_{L_L s} = n_{\overline{L}_L s} & \text{Heavy mass can be generated for vector-like
exotics}\\
& n_{L_R s} = n_{\overline{L}_R s} & \text{Heavy mass can be generated for vector-like
exotics}\\
\end{eqnarray*}
where the contraints which generate the heavy masses have been imposed 
in order to generate LRS models which contain no chiral exotics.

An initial classification run of $10^{9}$ distinct models was performed and 
the results are displayed in section \ref{results}. Due to a relative lack 
in abundance of three generation models a second run of $10^{11}$ distinct 
models was performed with the constraints on the vector-like chiral exotic 
states relaxed. Namely, the condition that $n_{1e} = n_{\overline{1}e}$ which 
arise from the Pati-Salam exotic sectors were relaxed, along with the 
conditions $n_{L_L s} = n_{\overline{L}_L s}$, $n_{L_R s} = n_{\overline{L}_R s}$ 
and $n_{1v} = n_{\overline{1}v}$ which arise from the LRS exotic sectors. 
We remark that whereas a $10^9$ run typically takes 2 days, a corresponding
$10^{11}$ run can take 28 weeks, which becomes prohibitive. 
The results of these two runs is presented and commented on in 
section \ref{results}.

\subsection{Top Quark Mass Coupling}
For a model to be phenomenologically viable, it must reproduce the spectrum of
the Standard Model while also reproducing the Standard Model interactions at
the low energy limit. Therefore, our analysis extends to classifying the number
of models which give the necessary conditions to include the top quark mass. In
this class of models the top quark mass coupling is
\begin{equation*}
\lambda_t Q^F u^{cF}h_u^B
\end{equation*}
where the superscripts $F$ and $B$ refer to the fermionic and bosonic
components of the associated superfield respectively. It has been shown that
the necessary conditions in order to have a top quark mass coupling can be
imposed by a straightforward general analytical method \cite{tqmc}. This
general method details, without loss of generality, that if $Q$, $u^c$ and
$h_u$ arise from the sectors $B^{(1)}_{pqrs}$, $B^{(2)}_{pqrs}$ and
$B^{(6)}_{pqrs} = B^{(3)}_{pqrs} + x$ respectively, there exists a top quark
mass coupling.

\subsection{Results}\label{results}
\begin{center}
\begin{figure}
\includegraphics[width=15cm]{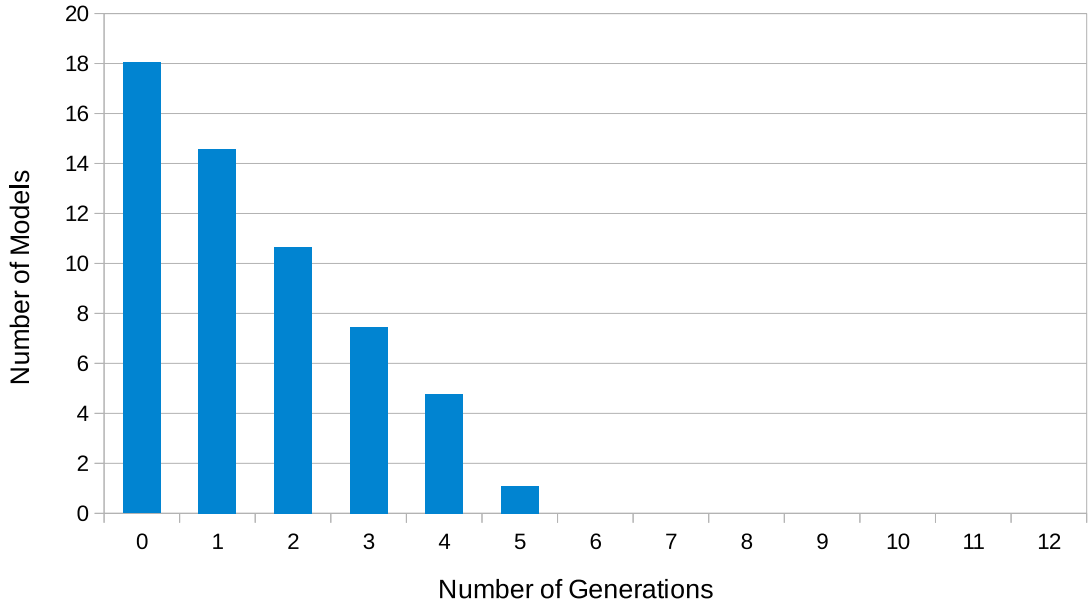}
\caption{\label{numberoffullgenanomfix} \emph{Natural logarithm of the number of
models against the number of generations ($n_g$) in a random sample of
$10^{9}$ GGSO configurations.}}
\end{figure}
\end{center}

\begin{center}
\begin{figure}
\includegraphics[width=15cm]{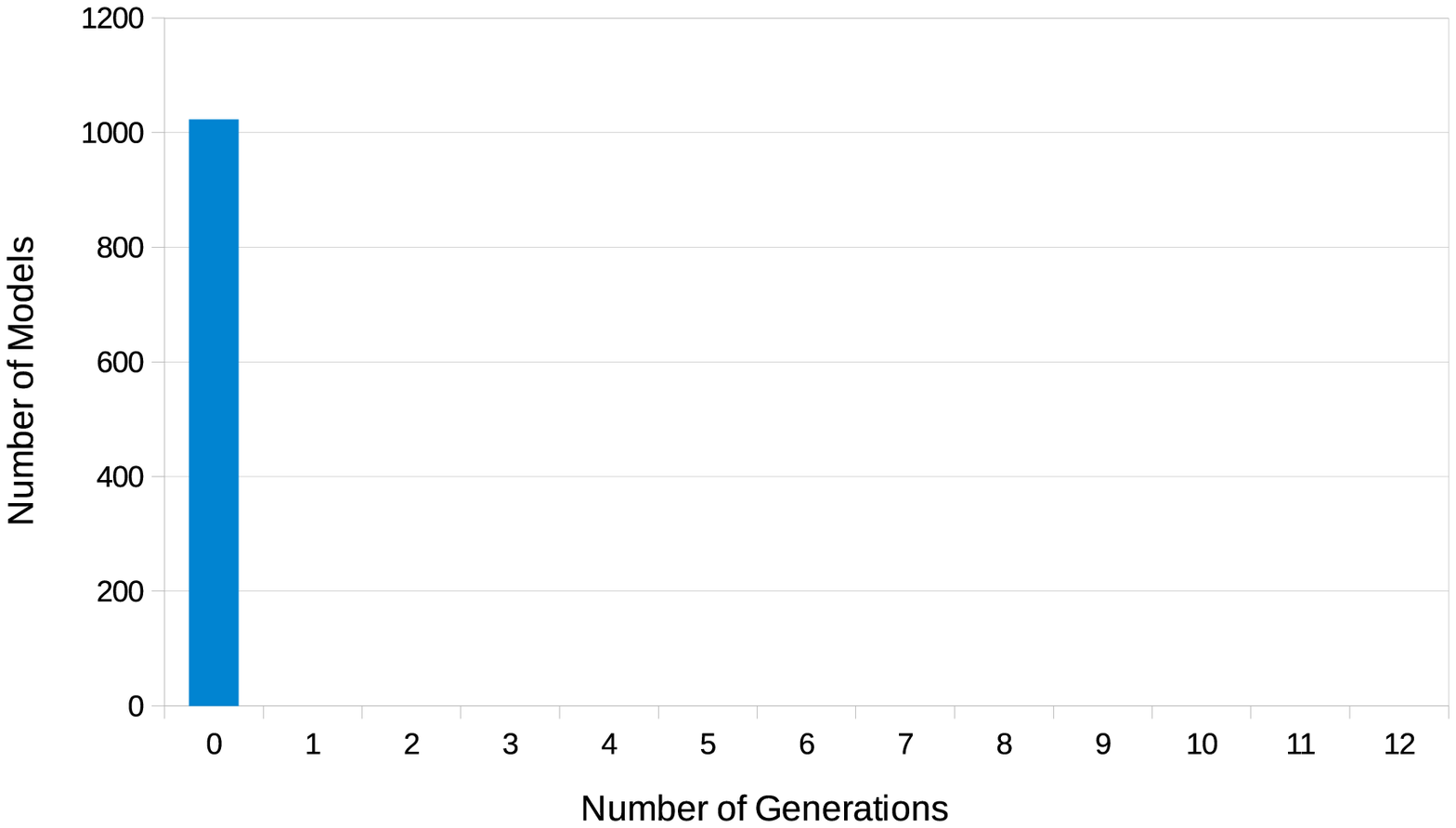}
\caption{\label{exophobicsanomfix} \emph{
Number of exophobic models against the number
of generations in a random sample of $10^{9}$ GGSO configurations. 
This figure should be contrasted with the corresponding figures in refs. 
\cite{acfkr} and \cite{frs}. }}
\end{figure}
\end{center}

\begin{center}
\begin{figure}
\includegraphics[width=15cm]{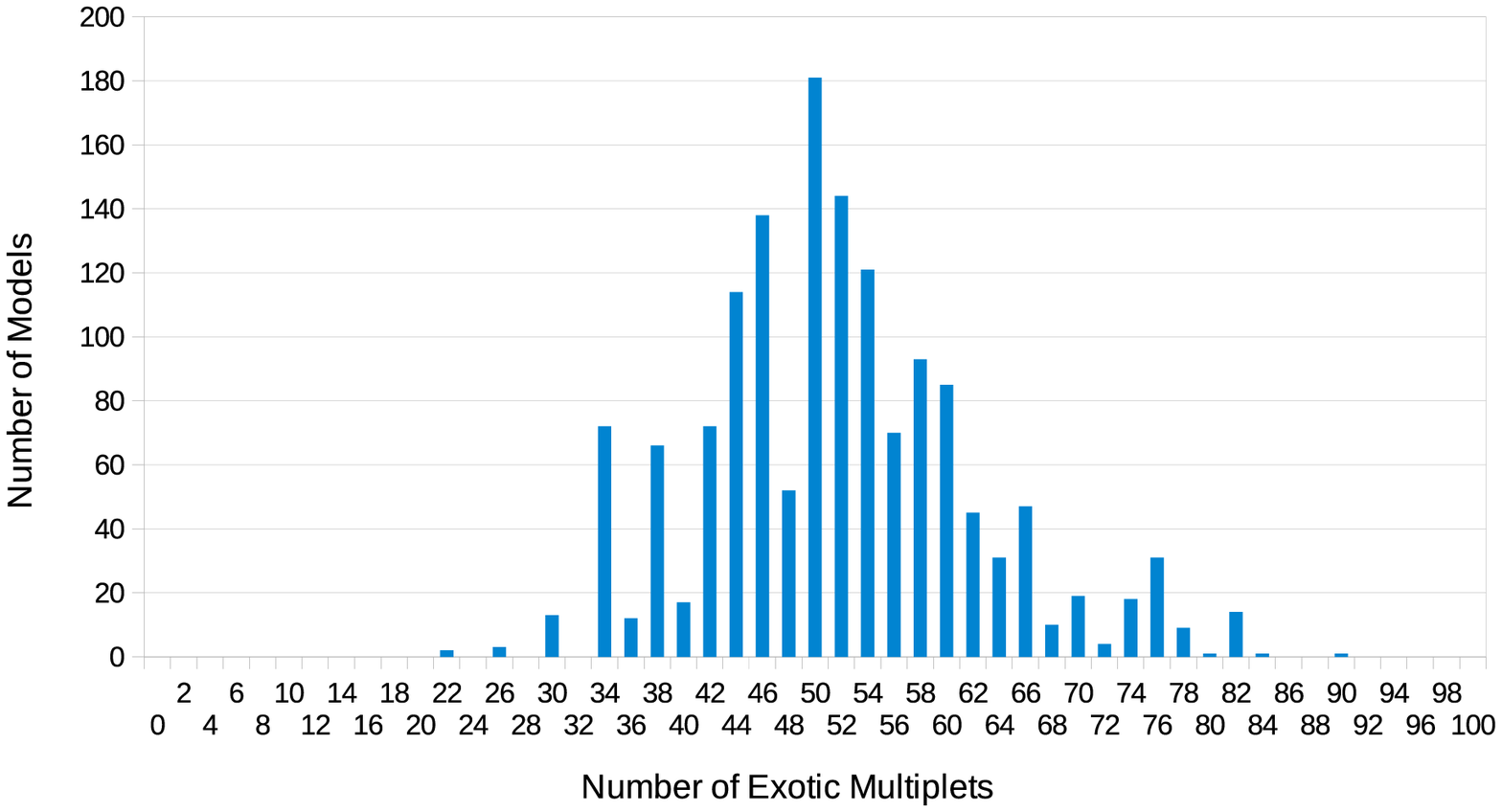}
\caption{\label{numberofexoticmultipletsanomfix} \emph{The number of
three generation models with no chiral exotic multiplets against the number of exotic multiplets in a random
sample of $10^{9}$ GGSO configurations.}}
\end{figure}
\end{center}

We now explore the space of the Left-Right Symmetric free fermionic heterotic
string vacua. The sample size used in the first classification was $10^{9}$ 
vacua
out of a possible total of $2^{66}$. Some of the results are presented in 
Figures \ref{numberoffullgenanomfix} - \ref{numberofexoticmultipletsanomfix} 
and table \ref{109table}.

In Figure \ref{numberoffullgenanomfix} the number of generations is presented 
against the natural logarithm of the number of models found. The results 
show the greatest number of models have zero generations and the number 
of models decreases as the number of generations increases. The maximum 
number of generations found was $n_g = 5$. Figure \ref{exophobicsanomfix} 
shows that only exophobic models with zero generations were found. 
Figure \ref{numberofexoticmultipletsanomfix} displays the number of 
three generation models with no chiral exotic multiplets found with 
respect to the total number of exotic multiplets they contain. 
The results show minimally exotic models to have 22 exotic 
multiplets while maximally exotic models have 90 exotic multiplets. 
The greatest number of models contained 50 exotic multiplets and 
the results show an approximately normal distribuition, skewed 
slightly to models containing more than 50 multiplets.
It can be seen in table \ref{109table} that $\approx 62.2\%$ of the 
non-enhanced models with complete families had no chiral exotics. 
The inclusion of the constraint demanding that the model must have three 
generations then drastically drops the probability of finding a viable model. 
The probability of finding a model which satisfies all these criteria 
is $1.49\times 10^{-6}$. Of these models, the probabilities that 
they contain no Higgs particles, only SM light Higgs particles 
or only heavy Higgs particles are $5.42\times 10^{-7}$, $9.39\times 10^{-7}$ 
and $7.00\times 10^{-9}$ respectively. Table \ref{109table} shows that 
requiring the model to contain both a light SM Higgs and a heavy 
Higgs yielded one model. Although this suggests models with interesting 
phenomenology exist, this result is not statistically significant and 
therefore does not allow meaningful conclusions to be drawn. 
This result also does not allow for any analysis involving further constraints.

Due to the lack of models with suitable phenomenology found during 
the $10^9$ sample, the sample size was increased to $10^{11}$ and 
some of the constraints were relaxed. Specifically, the constraints 
concerning the chiral exotic triplets in the models were 
included (\textit{i.e} $n_{3} = n_{\overline{3}}$ and 
$n_{3v} = n_{\overline{3}v}$), whereas the contraints concerning 
the vector-like chiral color--singlet exotics were omitted.
We note that relaxing these constraints entails that in some 
of the scanned models $U(1)_C$ is anomalous.

\begin{center}
\begin{figure}
\includegraphics[width=15cm]{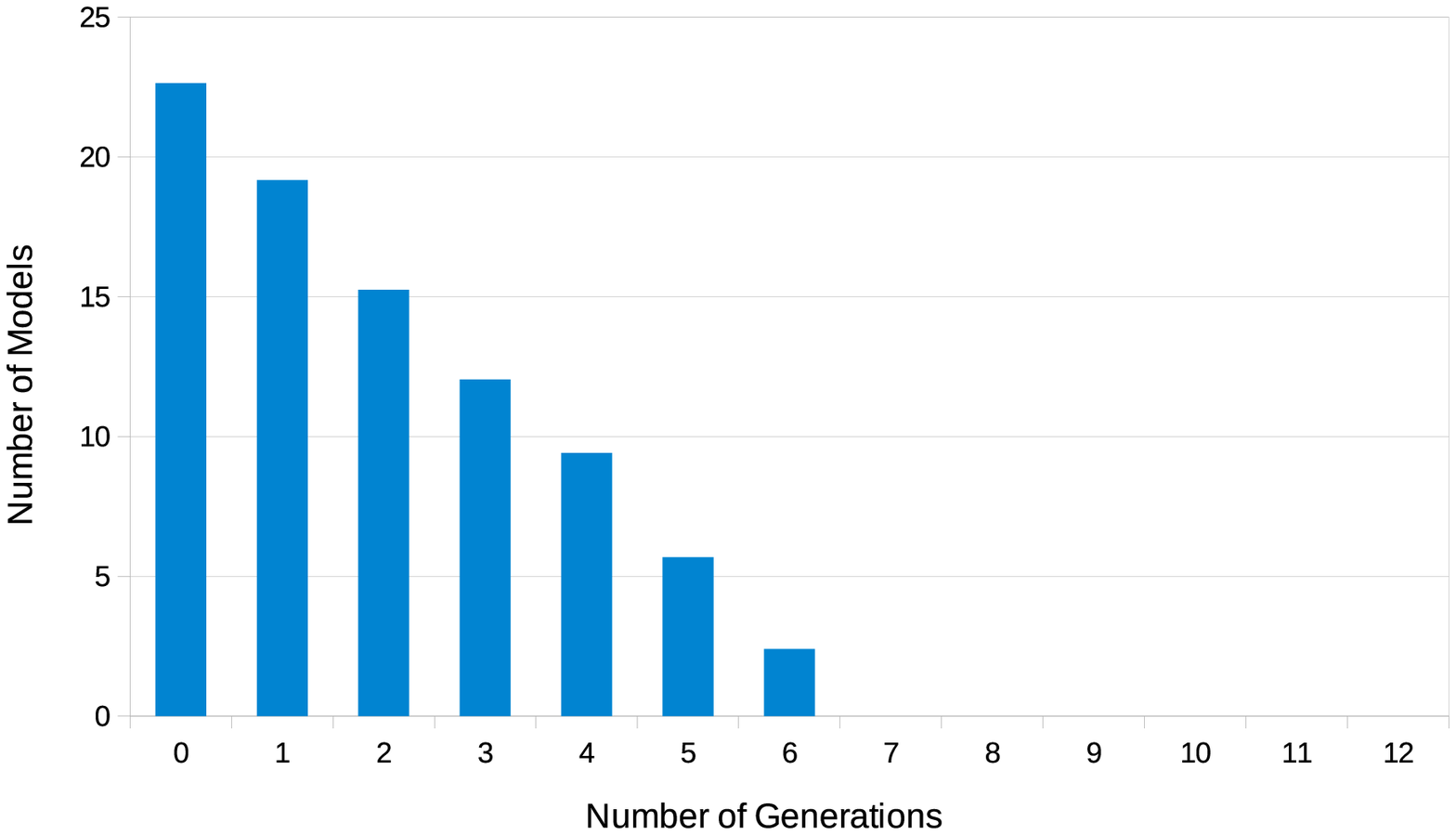}
\caption{\label{numberoffullgen} \emph{Natural logarithm of the number of
models against the number of generations ($n_g$) in a random sample of
$10^{11}$ GGSO configurations.}}
\end{figure}
\end{center}

\begin{center}
\begin{figure}
\includegraphics[width=15cm]{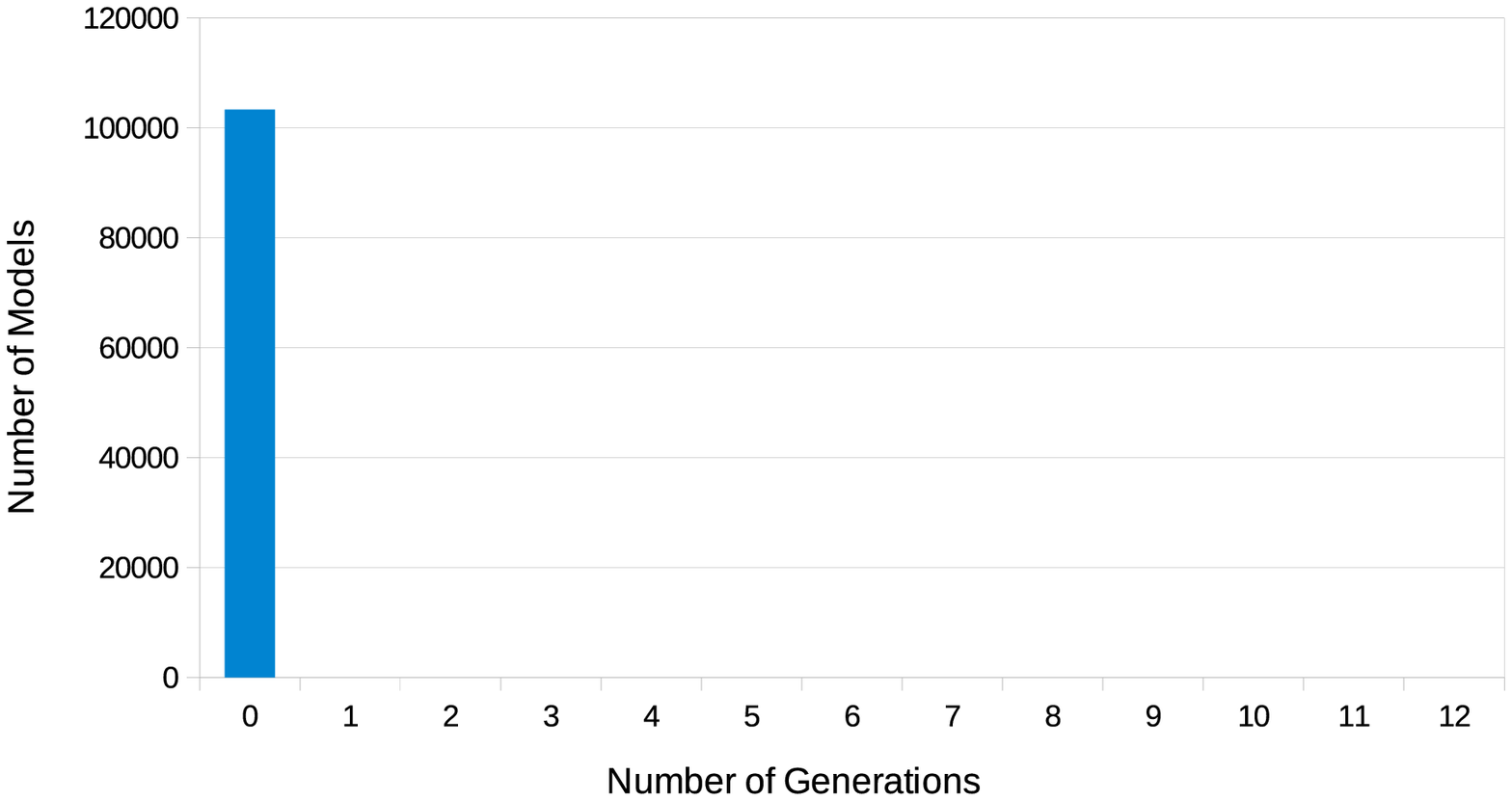}
\caption{\label{exophobics} \emph{Number of exophobic models against the number
of generations in a random sample of $10^{11}$ GGSO configurations. 
This figure should be contrasted with the corresponding figures in refs. 
\cite{acfkr} and \cite{frs}. }}
\end{figure}
\end{center}

\begin{center}
\begin{figure}
\includegraphics[width=15cm]{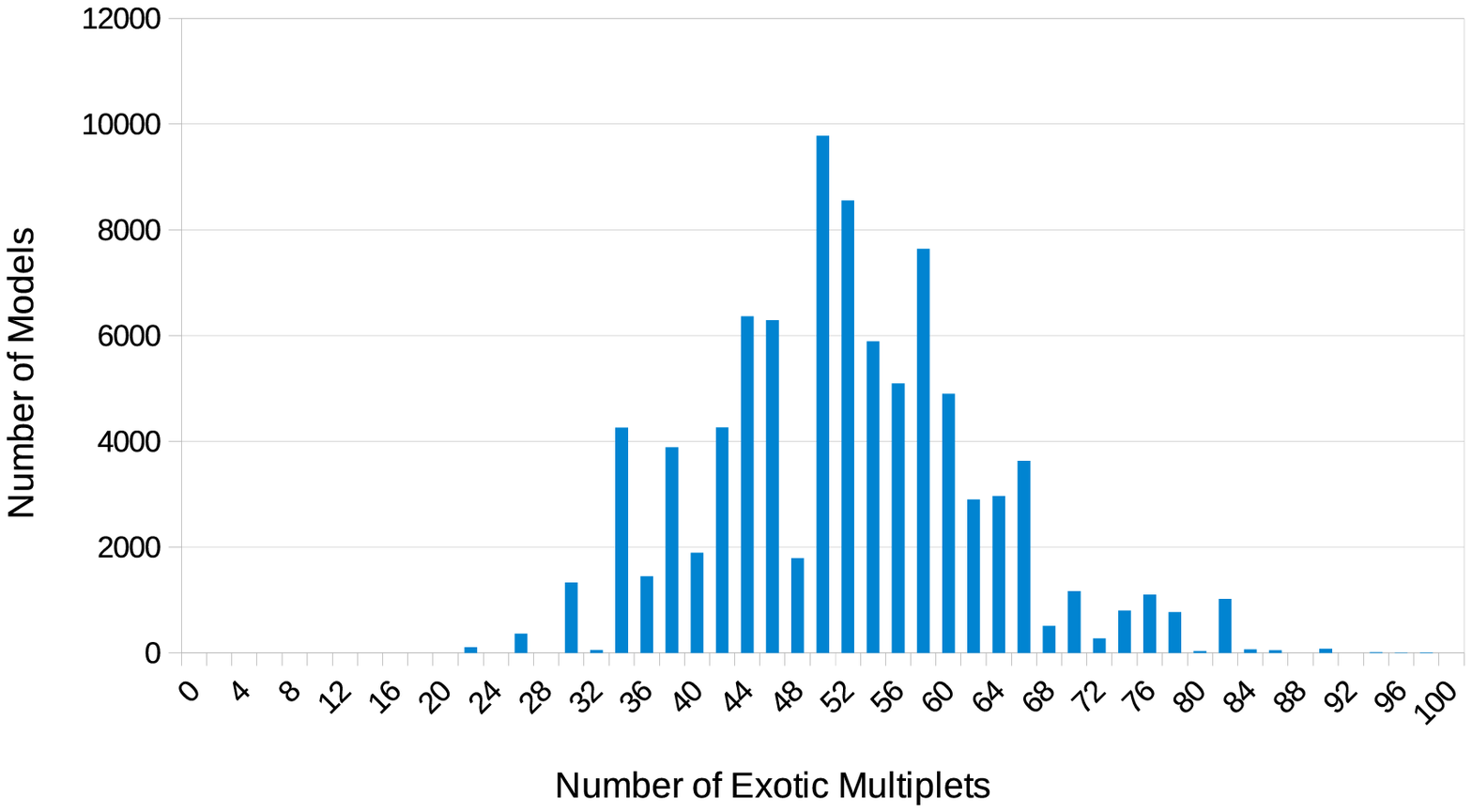}
\caption{\label{numberofexoticmultiplets} \emph{The number of three generation models with no chiral exotic triplets against the number of exotic multiplets in a random
sample of $10^{11}$ GGSO configurations.}}
\end{figure}
\end{center}

\footnotesize
\begin{table}
\begin{tabular}{|c|l|r|c|c|r|}
\hline
&Constraints & \parbox[c]{2.5cm}{Total models in sample}& Probability
&\parbox[c]{3cm}{ Estimated number of models in class}\\
\hline
 & No Constraints & $1000000000$ & $1$ & $7.38\times 10^{19}$ \\ \hline
(1)&{+ No Enhancements} & $708830165$ & $7.09\times 10^{-1}$ & $5.23\times
10^{19}$ \\  \hline
(2)& {+ Complete Families} & $70241057$ & $7.02\times 10^{-2}$ & $5.18\times
10^{18}$ \\ \hline
(3)&{+ No Chiral Exotics} & $43660665$ & $4.37\times 10^{-2}$ & $3.30\times
10^{18}$ \\  \hline
(4)&{+ Three Generations} & $1486$ & $1.49\times 10^{-6}$ &  $1.10\times
10^{14}$ \\  \hline
(5)&{+ SM Light Higgs}& $1$ & $1.00\times 10^{-9}$ & $7.38\times 10^{10}$ \\
&{+ \& Heavy Higgs}&&&  \\\hline
(6)&\parbox[c]{4cm}{+ Minimal Heavy Higgs} & 0 & $0$ &
$N/A$ \\
&{ \& Minimal SM Light Higgs}&&&  \\\hline
(7)&{+ Top Quark Mass Coupling } & 0 & $0$ & $N/A$ \\\hline
\end{tabular}
\caption{\label{109table} \emph{Statistics for the LRS models with respect
to phenomenological constraints for the sample of $10^9$ models.}}
\end{table}
\normalsize
The sample size was then increased to perform a classification on $10^{11}$ vacua
out of a possible total of $2^{66}$ and the program was run again. 
Some of the results are presented in
Figures \ref{numberoffullgen} - \ref{numberofexoticmultiplets} and Tables
\ref{exotics} - \ref{summarytable}.

In Figure \ref{numberoffullgen} the number of models versus the number of full
generations is displayed for the $10^{11}$ model run. The greatest 
number of models can be seen to have
zero generations and the number of models decreases as the number of
generations increases. This result is in accordance with the $10^9$ run 
and the previous results
of classifications \cite{acfkr, frs, SLM}. It can be seen that once the
number of generations is greater than six, there is an absence of models. This
result indicates that for this choice of basis vectors, models with $n_g \geq
7$ are either completely forbidden or are extremely unlikely in the total space
of model possibilities.

Figure \ref{exophobics} displays the number of exophobic models versus the
number of generations. Analogously to the $10^9$ classification run, 
the results show a relative abundance of zero generation
exophobic models but an absence of any exophobic models with $n_g \geq 1$. This
result leads to the conclusion that there are no three generation exophobic
models with a statistical frequency larger than $1 : 10^{11}$. It should
however be noted that the lack of exophobic models with $n_g \geq 1$ does not
suggest that exophobic Left-Right Symmetric models are completely forbidden,
only that for the choice of basis vectors used in this analysis none were found
with a reasonable statistical likelihood.

This result is in contrast to the case of the results of both the Pati-Salam and
Flipped $SU(5)$ classifications \cite{acfkr, frs}. In the Pati-Salam case,
exophobic models with $n_g = 0, \ldots, 6$ were found and where $n_g \geq 7$
exophobic models with an even number of generations were found with a notable
absence for $n_g = 14$. In the flipped $SU(5)$ case, exophobic models with an
even number of generations were found. While this means no three generation
exophobic models were found, the flipped $SU(5)$ case admits many more
exophobic models than the current LRS case.

In Figure \ref{numberofexoticmultiplets} the total number of three
generation models with matched number of colour triplets is displayed 
against the number of exotic fractionally
charged multiplets in a given three generation model. It can be
seen that the minimal number of exotic multiplets was again found 
to be 22, while the
maximally exotic models contained 98, an increase from the previous run. 
The results again show a roughly normal
distribution with a central peak at 50 exotic multiplets with a slight 
skew toward models where the number of exotic multiplets greater than 50. 
This result is
similar to what was found in the classification of Pati-Salam models
\cite{acfkr}, but in the case of the LRS the average number of exotic
multiplets is much higher. In the Pati-Salam case, there was a central peak at
18 exotic multiplets with maximally exotic models having 54 multiplets. This
result is, in general, to be expected as in the LRS models both Pati-Salam and
LRS exotic sectors exist, therefore there is the potential for many more exotic
states to enter the spectrum.

Table \ref{exotics} shows the number of non-enhanced three
generation models which have matched numbers of colour triplets with respect 
to the number of Pati-Salam, spinorial and
vectorial exotic multiplets. It can be seen that of the total number of models,
there were models found which contained no spinorial exotic multiplets. This is
also true in the case of vectorial exotic multiplets. However, no models were
found which were exophobic with respect to the Pati-Salam exotic multipets,
which is a leading reason for the lack of exophobic three generation models.
This result is in contrast to the results of the classification performed in
\cite{acfkr} as three generation exophobic Pati-Salam models were found.

Of the total models sampled, $\approx 61.1\%$ of the non-enhanced full
generation models were found to have matched numbers of colour triplets. 
This is a slight 
increase from the $10^9$ classification run due to the relaxing 
of some of the conditions as mentioned previously. This can be seen in table
 \ref{summarytable}. It should be noted that the probability of finding 
non-enhanced three generation models is actually lower in the $10^{11}$ 
classification run than in the $10^9$ case. This is expected to be a 
statistical fluctuation due to the random nature of the classification 
method. 
Further analysis on the effect of relaxing the requirement
that all color--singlet exotics are vector--like in three
generation models may be an interesting area of research. 
However, this is beyond the scope of this analysis and is left 
for future work.

If the constraint of having a top quark mass coupling is
 included, then of the total number of non-enhanced full generation models
 only $\approx 0.015\%$ were found to be viable. While three generation models
 with a top quark mass coupling were found, it can be seen from table
 \ref{summarytable} that their appearance was not found to be frequent, as the
 probability for finding such a model was found to be $4.0\times 10^{-11}$.

Of all the non-enhanced models with complete generations, $\approx 46.0 \%$
contained at least one light Higgs. This is much higher than for the case of
the heavy Higgs, where only $\approx 14.0\%$ of the total non--enhanced,
generation complete models contained at least one heavy Higgs. When
considering non--enhanced three generation models in which all exotic
colour triplet are vector--like, the number which had at least one 
Standard Model
Higgs is approximately $57.5\%$ and the number which had at least one heavy
Higgs is approximately $0.57\%$. 
Only $0.03\%$ of the non--enhanced three generation models with 
vector--like exotics contain both light and heavy Higgs multiplets.
Comparing with previous classifications, we note that in 
the three generation Pati-Salam models classified in
\cite{acfkr}, $7.9\%$ had a heavy Higgs and $81.0\%$ of these had a SM
Higgs. Whereas, in the flipped $SU(5)$ case \cite{frs}, the non--enhanced
and anomaly free three generation models had $\approx 95.7\%$ which contained a
SM Higgs and $\approx 6.3\%$ contained a heavy Higgs. In comparison, it can
be seen that the number of three generation free fermionic LRS models, 
free of $U(1)_C$ anomalies and enhancements,
which contain either Higgs is drastically lower. 
This outcome should nevertheless be compared with the case of the 
SU421 models in which no viable models can be constructed at all! 

\subsection{A Model of Notable Phenomenology}\label{notable}

The random classification method can be used to trawl models with
specified phenomenological properties.
Using the notation convention
\begin{equation}
C\binom{v_i}{v_j} = e^{i\pi(v_i | v_j)} 
\end{equation}
the model defined by the GGSO projection coefficients 
in eq. (\ref{exemplarymodel}) provides 
an example of a non--enhanced three generation model, 
with potentially viable phenomenology. 
\begin{equation}
(v_i | v_j ) = 
\bordermatrix{
~ & \mathds{1} & S & e_1 & e_2 & e_3 & e_4 & e_5 & e_6 & b_1 & b_2 & z_1 & z_2 & 
\alpha\cr
\mathds{1} & 1 & 1 & 0 & 0 & 0 & 1 & 0 & 1 & 1 & 1 & 0 & 1 & -\frac{1}{2} \cr 
S & 1 & 1 & 1 & 1 & 1 & 1 & 1 & 1 & 1 & 1 & 1 & 1 & 1 \cr
e_1 & 0 & 1 & 1 & 1 & 0 & 0 & 0 & 0 & 0 & 0 & 1 & 0 & 0 \cr
e_2 & 0 & 1 & 1 & 1 & 0 & 0 & 0 & 0 & 0 & 1 & 1 & 1 & 1 \cr
e_3 & 0 & 1 & 0 & 0 & 1 & 1 & 0 & 0 & 1 & 0 & 0 & 0 & 0 \cr
e_4 & 1 & 1 & 0 & 0 & 1 & 0 & 0 & 0 & 0 & 0 & 0 & 0 & 1 \cr
e_5 & 0 & 1 & 0 & 0 & 0 & 0 & 1 & 1 & 0 & 0 & 1 & 0 & 1 \cr
e_6 & 1 & 1 & 0 & 0 & 0 & 0 & 1 & 0 & 0 & 0 & 1 & 0 & 0 \cr
b_1 & 1 & 0 & 0 & 0 & 1 & 0 & 0 & 0 & 1 & 1 & 1 & 0 & 1 \cr
b_2 & 1 & 0 & 0 & 1 & 0 & 0 & 0 & 0 & 1 & 1 & 0 & 1 & 1 \cr
z_1 & 0 & 1 & 1 & 1 & 0 & 0 & 1 & 1 & 1 & 0 & 0 & 1 & 0 \cr
z_2 & 1 & 1 & 0 & 1 & 0 & 0 & 0 & 0 & 0 & 1 & 1 & 1 & 0 \cr
\alpha & 1 & 1 & 0 & 1 & 0 & 1 & 1 & 0 & 0 & 0 & 1 & 1 & 1}
\label{exemplarymodel}
\end{equation}
The observable matter sectors of this model produce three chiral generations, 
a minimal SM Higgs ($n_h = 1$) and a minimal heavy Higgs ($n_H = 1$). 
There exists colour triplets from the vectorial \textbf{10} representation of 
$SO(10)$ as $n_3 = 1$ and $n_{\overline{3}} = 1$, but as there are equal numbers 
of them heavy mass can be generated and there exists no anomaly in the 
LRS gauge group from these sectors. This model also contains no 
enhancements. The numbers defined in table \ref{phenonumbers} 
for the spinorial LRS exotic sectors of this model are as follows: 
$n_{L_L s} = n_{\overline{L}_L s} = 1$, $n_{L_R s} = n_{\overline{L}_R s} = 1$. 
The vectorial LRS exotics have the values $n_{3v} = n_{\overline{3}v} = 1$ 
and $n_{1v} = n_{\overline{1}v} = 5$. The Pati-Salam exotic states have the 
values $n_{L_L e} = 4$, $n_{L_R e} = 10$ and 
$n_{3e} = n_{\overline{3}e} = n_{1e} = n_{\overline{1}e} = 0$. 
The model therefore has no anomaly under the LRS gauge group, 
{\it i.e.} all the exotic states are in vector--like representations,
but does contain an anomaly under the $U(1)_2$ and $U(1)_3$ gauge groups. 
The model contains exotic multiplets and is therefore not exophobic. 
The model also admits a top quark mass coupling of order one. 

\begin{center}
\begin{table}
\begin{tabular}{|c|l|r|c|c|r|}
\hline
\parbox[c]{1.9cm}{\# Exotic Multiplets} & Pati-Salam & Spinorial & Vectorial\\
\hline
0 & 0 & 5536 & 1720\\
\hline
2 & 0 & 0 & 0 \\
\hline
4 & 0 & 20854 & 3215\\
\hline
6 & 0 & 0 & 0 \\
\hline
8 & 0 & 26727 & 19764\\
\hline
10& 0 & 0 & 0 \\
\hline
12& 319 & 19102 & 4272\\
\hline
14& 3030 & 0 & 0\\
\hline
16& 894 & 10616 & 19750\\
\hline
18& 15580 & 0 & 0\\
\hline
20& 18598 & 1648 & 2157\\
\hline
22& 13014 & 0 & 0\\
\hline
24& 8703 & 3796 & 18673\\
\hline
26& 15918 & 0 & 0\\
\hline
28& 3528 & 739 & 1532\\
\hline
30& 3386 & 0 & 0\\
\hline
32& 1797 & 169 &8093 \\
\hline
34& 2632 & 0 & 0\\
\hline
36& 1169 & 0 & 952\\
\hline
38& 398 & 0 & 0\\
\hline
40& 25 & 73 & 7209\\
\hline
42& 233 & 0 & 0\\
\hline
44& 0 & 0& 600\\
\hline
46& 35 &0 & 0\\
\hline
48& 0 & 0& 1212\\
\hline
50& 1 & 0& 0\\
\hline
52& 0& 0& 9\\
\hline
54&0 & 0& 0\\
\hline
56&0 & 0& 46\\
\hline
58& 0& 0& 0\\
\hline
60& 0& 0& 40\\
\hline
62& 0& 0& 0\\
\hline
64& 0& 0& 16\\
\hline
\end{tabular}
\caption{\label{exotics} \emph{The number of models is presented with respect
to the number of Pati-Salam, Spinorial and Vectorial exotic multiplets.}}
\end{table}
\end{center}

\section{Conclusion}\label{conclusion}
The Standard Model of particle physics provides viable parameterisation of all
subatomic observable data. Furthermore, the Standard Model may prevail
in providing such effective viable parameterisation
up to the GUT or Planck scales, where this necessarily breaks down
due to quantum gravity effects. If this is the avenue chosen by
nature, it is evident that further fundamental insight into the Standard
Model parameters can only be gained by embedding it in a theory of
quantum gravity. The synthesis of gravity and quantum mechanics
is not yet an accomplished feat. There are various approaches and from
a purely theoretical perspective all should be regarded on equal footing.
String theory is among these approaches. String theory, however, has
one key advantage. Its consistency conditions mandate the existence
of the gauge and matter structures that appear in the Standard Model.
Furthermore, these consistency conditions restrict the type and
enumeration of states that can appear in the construction. String theory
therefore provides the arena for the development of a phenomenological
approach to the synthesis of gravity and the gauge interactions.
Since the mid-eighties detailed quasi--realistic string models were
constructed. A particular class of phenomenological string vacua
are the $\mathds{Z}_2\times \mathds{Z}_2$ orbifold compactifications that were
constructed in the free fermionic formulation of the
heterotic string.

In this paper we extended the classification of phenomenological
free fermionic heterotic string vacua to models in which the
$SO(10)$--GUT group is broken to the Left--Right Symmetric (LRS),
$SU(3)_C\times U(1)_{B-L}\times SU(2)_L\times SU(2)_R$, subgroup.
NAHE--based \cite{nahe} LRS free fermionic models that utilise asymmetric
boundary conditions were constructed in ref. \cite{lrs}.
The free fermionic classification method adopted
herein utilise solely symmetric boundary conditions.
The class of LRS vacua gives rise to several distinct
features as compared to the FSU5, PS and SLM classes.
In all the cases we may denote a basis vector that breaks
the $SO(10)$ symmetry by $\alpha$. The difference between
the FSU5, PS and SLM models versus the LRS and the SU421 models
is that in the first case the vector $2\alpha$ does
not break the $SO(10)$ symmetry, whereas in the second
case it does. This distinction impacts the phenomenolgical
characteristics of the two classes. In the case of the
SU421 models the consequence is that it is not possible
at all to construct SU421 free fermionic models with complete
matter generations \cite{su421,421}. 
Thus, a class of models that has attractive
phenomenological features from a purely GUT--QFT perspective
\cite{wiseperez}, cannot be constructed as an heterotic--string
model, at least in the free fermionic formulation. By contrast,
viable LRS free fermionic models can be constructed and in some
abundance, as demonstrated in this paper. The key difference 
between the LRS and SU421 models is that in the later case
the $2\alpha$ projection chooses either the left-- or right--handed 
Standard Model representations, whereas in the former it does not.
In the case of the LRS models there is a remaining freedom 
in the charges of the $U(1)_j$ symmetries, $j=1,2,3$,
that produces opposite charges of the left-- and right--handed 
Standard Model representations. This is in marked contrast 
with the corresponding charges in the PS, FSU5 and SLM models, 
in which they are necessarily the same. It would be of interest
to explore how, and whether, these different cases are replicated
in terms of bundles on complex manifolds.

The class of LRS free fermionic models that we explored
herein differs from those of ref. \cite{lrs}. The difference
being that while the construction in ref. \cite{lrs}
does not admit an $x$--map \cite{xmap},
the free fermionic classification methodology
utilises this map. In this classification method
the $x$--map is used to generate the sectors that
produce the Standard Model electroweak Higgs
multiplets. The fact these LRS models do contain
the $x$--map, as well as the fact that the vector $2\alpha$
breaks the $SO(10)$ symmetry, results in the relative
scarcity of viable models in this class, compared to
the FSU5, PS and SLM cases. Additionally, it results in the
proliferation of exotic states in the LRS models as compared
to the other cases. A common feature of the LRS and SLM models
is that both cases contain two $SO(10)$ breaking basis vectors,
whereas the FSU5 and PS models contain a single one. This
results in the relative suppression in the LRS and SLM cases
of vacua with complete three families, as compared to the FSU5 and
PS cases. In ref. \cite{SLM}, for that reason,
the classification method was adapted to generate
randomly fertile $SO(10)$ cores, around which
complete SLM classification was performed. This was achieved
by identifying specific patterns in the $14\times 14$ matrix
of GGSO phases. Thus, we note that the utility of the
random generation method may have reached its limit, and
novel computer methods may be of benefit. Such tools may be 
particularly useful in analysis of non--supersymmetric 
string vacua \cite{nonsusy} and trying to uncover novel symmetries
that underlie the space of phenomenological
string compactifications \cite{moonshine}.

\footnotesize
\begin{table}
\begin{tabular}{|c|l|r|c|c|r|}
\hline
&Constraints & \parbox[c]{2.5cm}{Total models in sample}& Probability
&\parbox[c]{3cm}{ Estimated number of models in class}\\
\hline
 & No Constraints & $100000000000$ & $1$ & $7.38\times 10^{19}$ \\ \hline
(1)&{+ No Enhancements} & $70882805410$ & $7.09\times 10^{-1}$ & $5.23\times
10^{19}$ \\  \hline
(2)& {+ Complete Families} & $7023975614$ & $7.02\times 10^{-2} $ & $5.18\times
10^{18}$ \\ \hline
(3)&{+ No Chiral Exotic Triplets} & $4291254503$ & $4.29\times 10^{-2}$ & $3.17\times
10^{18}$ \\  \hline
(4)&{+ Three Generations} & $89260$ & $8.93\times 10^{-7}$ &  $6.59\times
10^{13}$ \\  \hline
(5)&{+ SM Light Higgs}& $29$ & $2.9\times 10^{-10}$ & $2.14\times 10^{10}$ \\
&{+ \& Heavy Higgs}&&&  \\\hline
(6)&\parbox[c]{4cm}{+ Minimal Heavy Higgs} & 22 & $2.2\times 10^{-10}$ &
$1.62\times 10^{10}$ \\
&{ \& Minimal SM Light Higgs}&&&  \\\hline
(7)&{+ Top Quark Mass Coupling } & 4 & $4.0\times 10^{-11}$ & $2.95\times
10^{9}$ \\\hline
\end{tabular}
\caption{\label{summarytable} \emph{Statistics for the LRS models with respect
to phenomenological constraints for $10^{11}$ models.}}
\end{table}

\newpage

\bibliographystyle{unsrt}

\end{document}